\documentclass[aps,prb,twocolumn,showpacs,nobibnotes]{revtex4}
\usepackage{graphics,graphicx,amsfonts,amsmath,amsbsy,amssymb,color}
\usepackage{subfig}
\usepackage{psfrag}  
\newcommand{\eigenk}[1] {\epsilon \left(\bfk_{#1} \right)}

\def \bfr {{\bf r}}

\def \bfi {{\bf i}}

\def \bfk {{\bf k}}

\def \bfk {{{\bf k}}}

\def \bfg {{\bf g}}
\def \bfq {{\bf q}}

\def \bfx {{\bf x}}

\def \beq {\begin{eqnarray}}
\def \eeq {\end{eqnarray}}

\def \iFCIQMC {{\mbox{\emph{i}-FCIQMC }}}
\def \iFCIQMCbracket {{\mbox{\emph{i}-FCIQMC}}}

\def \bfi {{\bf i}}

\def \ket {{\rangle}}
\def \bra {{\langle}}
\newcommand{\Hamil}{\hat{H}}

\newcommand{\potv}{\hat{v}_{12}}
\newcommand{\vft}[1]{v_{#1}}

\newcommand{\refeq}[1]{{Eq.~(\ref{#1})}}
\newcommand{\reffig}[1]{{Fig.~\ref{#1}}}
\newcommand{\refsec}[1]{{Sec.~\ref{#1}}}
\newcommand{\half}{\frac{1}{2}}

\begin{document}
\title{Convergence of many-body wavefunction expansions using a plane wave basis: from the homogeneous electron gas to the solid state}

\author{James~J.~Shepherd$^{(a)}$}
\email{js615@cam.ac.uk}
\author{Andreas~Gr\"{u}neis$^{(a)}$}
\author{George~H.~Booth$^{(a)}$}
\author{Georg~Kresse$^{(b)}$}
\author{Ali~Alavi$^{(a)}$}
\email{asa10@cam.ac.uk}
\affiliation{$^{(a)}$ University of Cambridge, Chemistry Department, Lensfield Road, Cambridge CB2 1EW, U. K.}
\affiliation{$^{(b)}$ University of Vienna, Faculty of Physics and Center for Computational Materials Science, Sensengasse 8/12, A-1090 Vienna, Austria}

\begin{abstract}
Using the finite simulation-cell homogeneous electron gas (HEG) as a model, we investigate the convergence of the correlation energy to the complete basis set (CBS) limit in methods utilising plane-wave wavefunction expansions. Simple analytic and numerical results from second-order M\o ller-Plesset theory (MP2) suggest a $1/M$ decay of the basis-set incompleteness error where $M$ is the number of plane waves used in the calculation, allowing for straightforward extrapolation to the CBS limit. As we shall show, the choice of basis set truncation when constructing many-electron wavefunctions is far from obvious, and here we propose several alternatives based on the momentum transfer vector, which greatly improve the rate of convergence. This is demonstrated for a variety of wavefunction methods, from MP2 to coupled-cluster doubles theory (CCD) and the random-phase approximation plus second-order screened exchange (RPA+SOSEX). Finite basis-set energies are presented for these methods and compared with exact benchmarks. A transformation can map the orbitals of a general solid state system onto the HEG plane wave basis and thereby allow application of these methods to more realistic physical problems.
\end{abstract}
\date{\today}
\maketitle

\section{Introduction}

The exact wavefunction for the $N$-particle non-relativistic electronic Schr\"odinger equation can be expressed as an expansion of Slater determinants which span a complete $N$-electron space in which the problem is posed. These Slater determinants, in turn, are comprised of the antisymmetrized products of spin orbitals, the set of which form a complete one-electron space. In general, however, neither the complete $N$-electron space, nor the complete one-particle space can be represented exactly and unavoidably we must make do with $M$ spin orbitals, and, at most, the corresponding $\binom{M}{N}$ determinants in the $N$-electron Fock space that these spin orbitals can construct.

Even within this finite set of determinants, it is extraordinarily difficult to construct exact solutions and in practice one has to resort to approximate theories which in quantum chemistry form the set of standard models\cite{Helgaker}. These range from the single Slater determinant used in Hartree-Fock theory to the variationally optimised linear combination of the full set of Slater determinants found by Full Configuration Interaction (FCI)\cite{Knowles1984,Olsen1988}. The coupled-cluster and many-body perturbation series form two distinct hierarchies. The ground state energy retrieved by FCI is the variationally lowest that can be achieved from this one-electron basis, within the wavefunction ans\"atze prescribed, and so is often termed the exact solution in this basis. 

However, the true solution to the Schr\"odinger equation can only be reached using FCI in the limit that the finite one-particle basis spans all of space which typically entails $M\rightarrow\infty$. Since this limit can never be reached in practice, schemes must be devised to find the behaviour of expectation values to allow for extrapolation to this limit. The complete basis set correlation energy, the difference between the HF and FCI energies in the limit of $M\rightarrow\infty$, is an important goal in \emph{ab initio} electronic-structure theory. Here, we will concentrate on the convergence of the correlation energy noting that the convergence of the Hartree-Fock energy and orbitals is generally well-understood and in the case of real systems can be obviated with pseudopotentials or carefully chosen atom centred basis sets\cite{Paier2009,GillanHF}. However, the convergence of the correlation energy in a plane wave basis set, which has substantial contributions from electron-electron cusps, has not been widely investigated.

In studies of molecular systems, the CBS correlation energy can be reasonably well approximated by extrapolation. In doing so, a certain functional form of the correlation energy is assumed, which can be rationalised by a partial wave analysis of the wavefunction around the electron-electron cusp. Most wavefunction based calculations of atoms and molecules employ correlation consistent Gaussian type orbital (GTO) basis sets, first developed by Dunning and coworkers, that show systematic behavior for many atoms and molecules\cite{Dunning1989,Dunning1992}, converging as $1/X^3$ where $X$ refers to the cardinal number of the basis set\cite{Kutzelnigg}. Since this cardinal number refers to a principal expansion, the number of orbitals ($M$) increases as $X^3$, and this convergence is equivalent to $1/M$.

The application of quantum chemical wavefunction-based methods to the solid state is a young and emerging field.\cite{Nolan2009,Paulus2006,Paulus2011,Scuseria2001,Schwerdtfeger2009,Shiozaki2010,Hirata2001,Usvyat2011,Maschio2011,Grueneis2011,Grueneis2010,Marsman2009,Paulus2012} Even within this body of work, most of the approaches have relied on a basis set expansion in periodic GTOs, where the wealth of knowledge on the convergence properties of these basis sets is well established from decades of studies in molecular calculations. Far less work has been undertaken on the convergence of determinantal wavefunction expansions in a plane wave basis, despite presenting a number of advantages when working in the solid state. By specifying a single cutoff parameter, an arbitrarily large set of linearly independent and intrinsically periodic basis functions can be produced, which require no optimisation, are free of basis-set superposition error and well describe the nature of delocalised electrons, which are particularly difficult for expansion in a more localised basis. 

Since wavefunction-based theories will inevitably be much more computationally expensive, it is imperative to develop methods in which the convergence with respect to the one-electron basis is as rapid as possible. Although complete basis set results using extrapolation procedures have been presented for systems in a plane wave basis\cite{Grueneis2011,Grueneis2010,Marsman2009}, a systematic analysis and rigorous justification for these schemes is still lacking. Furthermore, the question arises as to whether more efficient basis set truncations exist within the complete plane wave set, which allow for a more reliable extrapolation to the complete basis set limit. This paper aims at a rigorous investigation of different extrapolation methods for the homogeneous electron gas (HEG), which is taken to be the archetypal solid state model system, in order to extend the practicality of correlated wavefunction expansions in plane waves.

The limiting behaviour of basis set convergence is due to the inability of determintantal expansions to describe the features of the electron cusp and this is independent of the precise parameterisation of the wavefunction arising from the underlying method. This allows scaling relationships to hold across the whole hierarchy of standard models. As such, this paper examines the behaviour of basis set incompleteness error in plane waves on the correlation energy of a finite $N$-electron gas by use of second-order M\o ller-Plesset theory (MP2), where analysis can be directly performed and numerically verified to gain a preliminary understanding of this error. This is possible due to the MP2 correlation energy of a finite electron gas being well-defined in spite of the divergent behaviour of this energy at the thermodynamic limit. 

We show that a more natural interpretation of basis sets in momentum space can be found that relates to the momentum transfer vector. This discussion gives rise to a new type of basis set truncation that we can use to better eliminate basis set incompleteness error in MP2 and other theories. We then move away from the electron gas as a model system to show how these findings can be transferred back to real, solid-state systems. We hope that this provides the first thorough analysis of basis set incompleteness in calculations where a plane wave basis set is used and will allow for extrapolations to the CBS limit to be found, both more reliably and more efficiently.

We note that extrapolation is not the only method by which basis set incompleteness error can be removed. It is now increasingly common practice in molecular quantum chemistry to use corrections based on including explicit functions of the inter-electronic distance into the wavefunction\cite{Shiozaki2010,Tew2012}. Furthermore, there have been significant advances in applying transcorrelated methods directly to the homogeneous electron gas\cite{BoysHandy,Ochi,Umezawa,Sakuma}. Diffusion Monte Carlo, which is in general not particularly sensitive to basis set, has also been incredibly successful in describing ground-state energies and properties for the HEG\cite{CeperleyAlder1980,Foulkes2001,OB,OB2,Holzmann2011,Gurtubay2010,Kwon1998,Rios,Needs,Drummond2009,Holzmann2009,Huotari,Drummond2009b}. Nonetheless, we believe that simple complete basis set extrapolation techniques would enable reliable benchmarks to be obtained for the future development of wavefunction techniques in periodic systems.

\section{An analysis of Plane Wave Basis Set Incompleteness Error}

In this section we will use the archetypal model solid state system, the homogeneous electron gas, to better-understand basis set incompleteness in plane waves. We will introduce the HEG Hamiltonian and show how MP2 theory can be applied to produce an analytic expression for the correlation energy approaching the complete basis set limit, which we verify numerically. Although it is well-known that the correlation energy arising from MP2 theory diverges in the thermodynamic limit due to long-wavelength excitations as the band gap closes\cite{GellMann}, the qualitative cusp behaviour as inter-electronic distance goes to zero is inherently captured by short-wavelength excitations\cite{Kimball}. As such, using MP2 as a model theory for correlation provides a good starting-point for our discussion of basis set incompleteness error\footnote{We note that other methods have also used MP2 theory as their starting-point for extrapolations of the correlation energy in molecular systems\cite{Ayala,Iyengar}.}.

\subsection{Using the electron gas as a model system}

%
The $N$-electron HEG simulation-cell Hamiltonian can be written:
\begin{equation}
\Hamil=\sum_\alpha -\half \nabla_\alpha^2 + \sum_{\alpha\neq \beta} \half \hat{v}_{\alpha\beta} + \half N v_\text{M}
\label{sim_cell_H}
\end{equation}
where $\alpha$ and $\beta$ are electron indices and the two-electron operator $\hat{v}_{\alpha\beta}$ is:
\begin{equation}
\hat{v}_{\alpha\beta}= \frac{1}{\Omega} \sum_\bfq v_\bfq e^{i \bfq \cdot \left( \bfr_{\alpha} - \bfr_\beta \right)} \quad ; \quad
v_\bfq = \left\{
\begin{array}{ll}
\frac{4\pi}{\bfq^2}, & \bfq\neq\bf{0} \\
0, & \mbox{\bfq=\bf{0}}
\end{array}
\right. 
\end{equation}
$v_\text{M}$ is the Madelung term, which represents contributions to the one-particle energy from interactions between a point charge and its own images and a neutralising background, and $\Omega$ is the real-space simulation cell volume. Together, all $\hat{v}_{\alpha\beta}$ and $v_\text{M}$ form what is termed the Ewald interaction\cite{Ewald,Fraser1996,Drummond2008}. Hartree atomic units (a.u.) are used throughout and energies quoted are total correlation energies for the system considered unless otherwise stated.

The one-electron basis set is taken to be plane waves,
\begin{equation}
\psi_{j}  (\bfx) \equiv \psi_{j}  (\bfr , \sigma) =\sqrt{\frac{1}{\Omega}}~e^{i \bfk_j \cdot \bfr} ~\delta_{\sigma_j,\sigma},
\end{equation}
where the wavevectors $\bfk_j$ are chosen to correspond to the reciprocal lattice vectors of a real-space cubic cell of length $L$, 
\begin{equation}
\bfk=\frac{2\pi}{L} \left(n,m,l\right),
\end{equation}
where $n$,$m$ and $l$ are integers and $\Omega=L^3$ is the real-space unit cell volume of a cubic cell. 

In this basis the HEG Fock matrix, is diagonal, and the Hartree-Fock determinant is the normalised, antisymmetrized product of $N$ plane waves with the lowest kinetic energy,
\begin{equation}
D_{\bf 0} = \mathcal{A} \left[ \psi_i(\bfx_1) \psi_j(\bfx_2) ... \psi_k(\bfx_N) \right]
\end{equation}
with the energy,
\begin{equation}
\bra D_{\bf 0} | \Hamil | D_{\bf 0} \ket= \frac{1}{2} \sum_i^N \bfk_i^2  - \frac{1}{\Omega} \sum_i^N \sum_{j>i}^N \frac{4 \pi}{\left| \bfk_i-\bfk_j\right|^2} + \half N v_{\text{M}},
\label{eq5}
\end{equation}
where the removal of the $\bfq=\bf 0$ term in the two-electron operator has removed the two-electron Coulomb term, corresponding physically to the cancellation of the classical interaction between the electrons and the interaction between the electrons and the neutralising background. The remaining terms in \refeq{eq5} are the kinetic energy, the exchange energy and the Madelung energy.

\subsection{Convergence of finite basis MP2 calculations}
{\label{sec1}}

\begin{figure}
\begin{centering}

\subfloat[Correlation energies retrieved as a function of $E_k^{-\frac{3}{2}}$. Energy in scaled units of $\left(\frac{2 \pi}{L}\right)^2$~a.u.]{
\psfrag{KKK}[l][l][1.0][0]{$E_k$-cutoff} 

\psfrag{AAA}[l][l][1.0][0]{CBS $\pm$ 1\%} 
\psfrag{BBB}[l][l][1.0][0]{CBS result} 
\psfrag{XXX}[][][1.0][0]{(Scaled energies)$^{-\frac{3}{2}}$}
\psfrag{YYY}[][][1.0][0]{Correlation energy / a.u.} 

\includegraphics[width=0.45\textwidth]{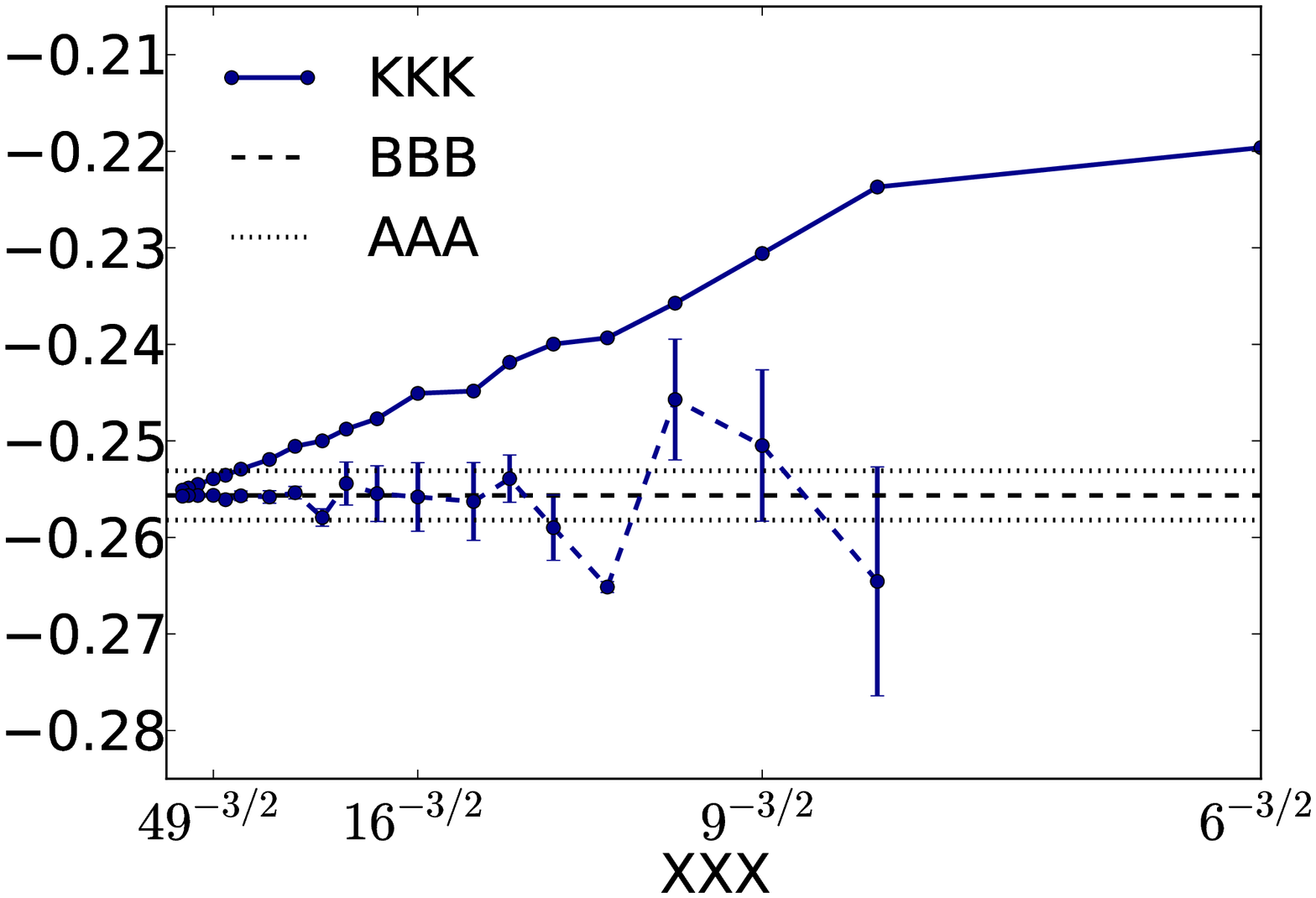}
\label{Ekgraph}

}

\subfloat[Correlation energies retrieved as a function of $M^{-1}$, where $M$ is the number of spin orbitals used.]{

\psfrag{KKK}[l][l][1.0][0]{$E_k$-cutoff} 

\psfrag{SSS}[][][1.0][0]{$162^{-1}$} 
\psfrag{SSS2}[][][1.0][0]{$246^{-1}$} 
\psfrag{SSS3}[][][1.0][0]{$514^{-1}$} 
\psfrag{SSS4}[][][1.0][0]{$2838^{-1}$} 

\psfrag{AAA}[l][l][1.0][0]{CBS $\pm$ 1\%} 
\psfrag{BBB}[l][l][1.0][0]{CBS result} 
\psfrag{XXX}[][][1.0][0]{$M^{-1}$}
\psfrag{YYY}[][][1.0][0]{Correlation energy / a.u.} 

\includegraphics[width=0.45\textwidth]{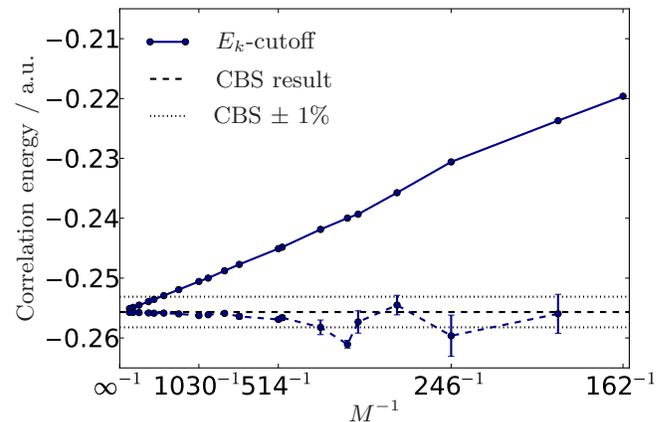}
\label{Mextrapgraph}

}

\caption{MP2 correlation energies for the 14 electron gas at $r_s=5.0$~a.u. retrieved as a function of $E_k^{-\frac{3}{2}}$ and $M^{-1}$ (where $M$ is the number of spin orbitals) tend towards a linear relationship as the complete basis set limit is approached. In each plot, the dotted lines refers to CBS limits for each basis set size, which are obtained by a linear extrapolation of this point and the previous three points (sometimes not visible on the graph). Using these extrapolated estimates, it can be seen that the $M^{-1}$ power-law is smoother due to fewer finite size effects, this is due to $M$ being a more appropriate variable to consider how much correlation energy the basis set retrieves. For this system, $M=1030$ corresponds to a kinetic energy cutoff of 1.3077 a.u. or 35.59 eV, which changes with both $N$ and $r_s$ for the HEG.}
\label{Fig1}

\end{centering}
\end{figure}

M\o ller-Plesset (MP) theory attempts to find the correlation energy of a system by treating the full electron-electron interaction perturbatively within Rayleigh-Schr\"odinger perturbation theory\cite{MP2paper}. Taking the zeroth-order Hamiltonian as the sum over Fock operators and the Hartree-Fock solutions as the zeroth-order wavefunctions, the first order energy is the Hartree-Fock energy. This makes the second-order term (MP2) the leading contribution to the correlation energy of the problem.

The MP2 correlation energy can therefore be expressed,
\begin{equation}
E_\text{MP2} = \sum_{\bfi \neq {\bf 0}} \frac{ | \bra D_\bfi | \Hamil^\prime | D_{\bf 0} \ket |^2 } {E_{\bf 0} - E_\bfi},
\end{equation}
where $\Hamil^\prime$ is the fluctuation operator defined as the difference between the Hamiltonian and the sum over the Fock operators. The zeroth-order wavefunctions $D_\bfi$ are the up to $N$-fold excitations of the Hartree-Fock determinants into a complete, typically infinite, basis. Truncating the basis set at some $M$ plane waves, these determinants are now the $\mathcal{O} \left[ \binom{M}{N} \right]$ rearrangements of $N$ electrons in $M$ spin orbitals. Since $\Hamil^\prime$ contains at most two-electron operators, only the $\mathcal{O} \left[ N^2 M \right]$ doubly excited determinants of $D_{\bf 0}$ make a contribution to this energy. Single excitations of the reference are not coupled to the reference due to Brillouin's theorem but also because, in the HEG, a single excitation necessarily forms a many-particle state of a different total momentum. Finally, the zeroth-order energies $E_\bfi$ are sums over the constituent orbital energies $\epsilon_i$, 
\begin{equation}
\begin{split}
&\epsilon_i= \half \bfk_i^2 -\sum_{\substack{j \in \text{occ} \\ j \neq i} } \bra i j | \potv | j i \ket - \half v_{\text{M}} \\
&\epsilon_a= \half \bfk_a^2 -\sum_{\substack{j \in \text{occ}} } \bra a j | \potv | j a \ket
\end{split}
\end{equation}
where in these equations, $i$ refers to any occupied orbital and $a$ refers to the virtual orbitals. The two-electron integrals can in general be evaluated as:
\begin{equation}
\begin{split}
\bra &i j | \potv | a b \ket = \delta_{\sigma_i,\sigma_a} \delta_{\sigma_j,\sigma_b}  \\
&\iint d\bfr_1 d\bfr_2 \psi_{i}  (\bfr_1) ^{\star} \psi_{j}  (\bfr_2) ^{\star} \potv \left( \bfr_1,\bfr_2 \right) \psi_{a}  (\bfr_1) \psi_{b}  (\bfr_2). \\
\end{split}
\end{equation}
These equations include an exchange energy explicitly, and in the thermodynamic limit tend towards the well-known form\cite{MartinElecStructure}
\begin{equation}
\epsilon_k = \half k^2 + \frac{k_F}{\pi} f\left(x \right)
\end{equation}
where $x=k/k_F$ and 
\begin{equation}
f\left(x \right)=\left( 1+\frac{1-x^2}{2x} \text{ln} \left| \frac{1+x}{1-x} \right| \right).
\end{equation}
This allows the MP2 energy to be re-written as,
\begin{equation}
E_\text{MP2} = 
\frac{1}{4}
\sum_{\substack{ij \in \text{occ} \\ ab \in \text{virt}}}
\frac{ | \bra i   j | \potv | a  b \ket - \bra i  j  | \potv | b  a \ket |^2 }{\epsilon_i + \epsilon_j-\epsilon_a - \epsilon_b},
\label{MP2-mastereq}
\end{equation}
where indices $i$,$j$,$a$ and $b$ are spin orbitals.

This can be solved directly for the non-interacting reference in the limit of both an infinite number of electrons and an infinite virtual k-space\cite{Onsager}. However, the limit of a finite number of electrons is dependent on the form of potential $\potv$ and the shape of the real space unit cell. Furthermore, it is typical to use a finite basis set to describe the virtual manifold, which can be achieved in the plane-wave basis with a choice of kinetic energy cutoff, $E_k=\half k_c^2$ such that,
\begin{equation}
\sum_{ij \in \text{occ}} \rightarrow \sum_{\sigma_i~\sigma_j} \sum_{\substack{0 \leq k_i \leq k_f \\ 0 \leq k_j \leq k_f  }} \quad ; \quad
\sum_{ab \in \text{virt}} \rightarrow \sum_{\sigma_i~\sigma_j}  \sum_{\substack{k_f < k_a \leq k_c \\ k_f <  k_b \leq k_c }}
\label{EQ:1}
\end{equation}
where $k_i=|\bfk_i|$ etc. and sums over spins have been written explicitly. Using this substitution, \refeq{MP2-mastereq} can be re-cast in a finite basis,
\begin{equation}
\begin{split}
E_\text{MP2} &= 
\sum_{\substack{0 \leq k_i \leq k_f \\ 0 \leq k_j \leq k_f  }}
\sum_{\substack{k_f < k_a \leq k_c \\ k_f < k_b \leq k_c }}
\frac{ 2 | \bra \bfk_i \bfk_j | \potv | \bfk_a \bfk_b \ket  |^2 }{ \eigenk{i} + \eigenk{j} -\eigenk{a}- \eigenk{b} } \\
&- 
\sum_{\substack{0 \leq k_i \leq k_f \\ 0 \leq k_b \leq k_f  }}
\sum_{\substack{k_f < k_a \leq k_c \\ k_f <  k_b \leq k_c }}
\frac{\bra \bfk_i \bfk_j | \potv | \bfk_a \bfk_b \ket \bra \bfk_i \bfk_j | \potv | \bfk_b \bfk_a \ket }{ \eigenk{i} + \eigenk{j} -\eigenk{a}- \eigenk{b}},
\end{split}
\end{equation}
where the sums over spins have been taken leaving a spin-free expression. The two sets of terms are referred to as direct and exchange-like terms respectively.

Defining $\vft{ \bfq }$ as the $\bfq$ Fourier component of the potential, the four index integrals can be evaluated, 
\begin{equation}
\bra \bfk_i \bfk_j | \potv | \bfk_a \bfk_b \ket = \vft{ \bfk_i-\bfk_a  } \delta_{\bfk_i-\bfk_a ,\bfk_b-\bfk_j }
\end{equation}
yielding,
\begin{equation}
\begin{split}
E_\text{MP2} = 
\sum_{\substack{0 \leq k_i \leq k_f \\ 0 \leq k_j \leq k_f }}
\sum_{\substack{k_f < k_a \leq k_c \\ k_f <  k_b \leq k_c }}
\delta_{\bfk_i-\bfk_a ,\bfk_b-\bfk_j } \frac{  2 ~ \vft{ \bfk_i-\bfk_a  }  ^2   }{\Delta \epsilon_{ijab}}  \\
-
\sum_{\substack{0 \leq k_i \leq k_f \\ 0 \leq k_j \leq k_f }}
\sum_{\substack{k_f < k_a \leq k_c \\ k_f <  k_b \leq k_c }}
\delta_{\bfk_i-\bfk_a ,\bfk_b-\bfk_j } \frac{  \vft{ \bfk_i-\bfk_a} \vft{\bfk_j-\bfk_a}   }{\Delta \epsilon_{ijab}}, 
\label{K3CUT}
\end{split}
\end{equation}
where $\Delta \epsilon_{ijab}$ is the difference between eigenvalues and depends on the four indices. Values of $\bfk_b$ in this representation are constrained to obey momentum conservation,
\begin{equation}
\bfk_i + \bfk_j = \bfk_a + \bfk_b,
\end{equation}
due to $\delta_{\bfk_i-\bfk_a ,\bfk_b-\bfk_j }$, and, therefore, the sum over $\bfk_b$ makes at most one contribution for every $\bfk_i$, $\bfk_j$ and $\bfk_a$. This is the formulation of the MP2 energy that we will refer to as the $E_k$-cutoff scheme.

The question we now seek to address is: how does the correlation energy captured by MP2 increase with the energy cutoff of the basis set on approach to the complete basis set limit? Since this question has not been addressed for plane wave basis sets, it is appropriate to conduct a simple analysis as follows.

We seek an expression for the error of a finite calculation conducted at a kinetic energy cutoff $E_k=\half k_c^2$,
\begin{equation}
\Delta E_\text{MP2} (k_c) = E_\text{MP2} (\infty) - E_\text{MP2} (k_c),
\end{equation}
where the $E_\text{MP2} (k_c)$ is the finite-basis MP2 energy given in \refeq{K3CUT}. This can be evaluated by changing the limits on the sums, such that:
\begin{equation}
\begin{split}
\Delta E_\text{MP2} (k_c) =
\sum_{\substack{0 \leq k_i \leq k_f \\ 0 \leq k_j \leq k_f }}
\sum_{\substack{k_a > k_c  \\ k_b >  k_c }}
\delta_{\bfk_i-\bfk_a ,\bfk_b-\bfk_j } \frac{  2 ~ \vft{ \bfk_i-\bfk_a  }  ^2   }{\Delta \epsilon_{ijab}}  \\
-
\sum_{\substack{0 \leq k_i \leq k_f \\ 0 \leq k_j \leq k_f }}
\sum_{\substack{k_a > k_c  \\ k_b >  k_c }}
\delta_{\bfk_i-\bfk_a ,\bfk_b-\bfk_j } \frac{  \vft{ \bfk_i-\bfk_a} \vft{\bfk_j-\bfk_a}   }{\Delta \epsilon_{ijab}} 
\end{split}
\end{equation}

It is possible to simplify this expression in the high basis set limit. The orbital energies become dominated by high-energy kinetic energy contributions, whereupon $\Delta \epsilon_{ijab} \propto k_a^2$. As $k_a \gg k_i$ and $k_a \gg k_j$, the numerator tends towards a behaviour of $1/k_a^4$. In this limit, the summation of $\bfk_i$ and $\bfk_j$ yields a constant factor, and the Kronecker delta reduces the double-sum over virtual orbitals to a single sum.

This leads to a leading-order expression of,
\begin{equation}
\Delta E_\text{MP2} (k_c) \propto
\sum_{\substack{k_a > k_c}}
\frac{1}{ k_a^6}
\end{equation}
where the sum can replaced by a spherically symmetric integral and evaluated as,
\begin{equation}
\begin{split}
\Delta E_\text{MP2} (k_c) & \propto \int_{k_c}^\infty \text{d} k_a~\frac{1}{k_a^6}~ k_a^2\\
& \propto \frac{1}{k_c^3}.
\label{INTEGRAL}
\end{split}
\end{equation}

This is equivalent to $E_k^{-\frac{3}{2}}$, due to the definition that $E_k=\half k_c^2$, or $M^{-1}$, where $M$ is the number of k-points contained within the sphere defined by $k_c$. In passing, we note that this is the behavior that is also found for the correction to the energy in the random phase approximation\cite{Harl2008}.

Figure \ref{Fig1} shows numerical verification of this relationship using $r_s=5.0$~a.u., a typical $r_s$ of real materials. In \reffig{Ekgraph}, a relatively rapid tendency to follow a $E_k^{-\frac{3}{2}}$ power-law is found. Extrapolated results at each basis set (using this basis set size and the previous three basis set sizes) show rapid convergence to the infinite basis set result, although this tendency is not smooth due to shell-filling effects. When instead a $M^{-1}$ power-law extrapolation is used, as in \reffig{Mextrapgraph}, this convergence is somewhat smoother and better behaved for small basis set sizes (when the difference between the two power-laws is more pronounced). 

\section{Momentum transfer vector cutoff schemes}
\label{MTVCS}

\begin{figure*}
  \centering
  \subfloat[This is a diagram of the \textit{local} $E_g$-cutoff. Using a simple momentum transfer cutoff scheme (\refeq{eq:G1}) makes the excitation space (set of possible virtual orbitals to be excited into), $\bfk_i$,$\bfk_j$ $\rightarrow$ \{$\bfk_i+\bfg$\},\{$\bfk_j+\bfg^\prime$\}, dependent on $\bfk_i$ and $\bfk_j$ when they are not at the $\Gamma$-point. This implies that the sets \{$\bfk_i+\bfg$\} and \{$\bfk_j+\bfg^\prime$\} are not the same.]{\label{fig:G1}\includegraphics[width=0.4\textwidth]{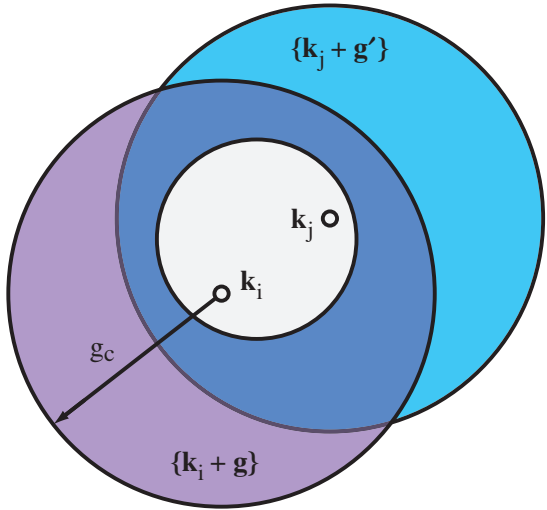}} \quad\quad\quad
  \subfloat[This is a diagram of the \textit{local} $E_g$-cutoff. Comparison between a specific excitation $\bfk_i$,$\bfk_j$ $\rightarrow$ $\bfk_a$,$\bfk_b$ and $\bfk_i$,$\bfk_j$ $\rightarrow$ $\bfk_b$,$\bfk_a$. These differ only in the permutation of the hole states (or, equivalently, the electron states). In the case of the \textit{local} $E_g$-cutoff cutoff scheme (\refeq{eq:G1}), one term (solid line) is allowed and the other term (dashed line) is disallowed.]{\label{fig:exG1}\includegraphics[width=0.4\textwidth]{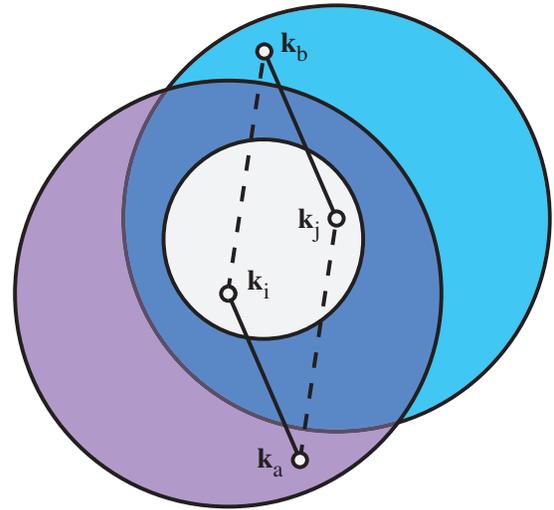}}

\subfloat[This is an illustration of the \textit{intersection} $E_g$-cutoff. One solution to the problem illustrated in \reffig{fig:exG1} is to only allow excitations to the region of k-space formed by the overlap of the two regions in \reffig{fig:G1}. Now both terms are either disallowed (as shown here) or allowed and permutational symmetry is restored.]{\label{fig:G2}\includegraphics[width=0.4\textwidth]{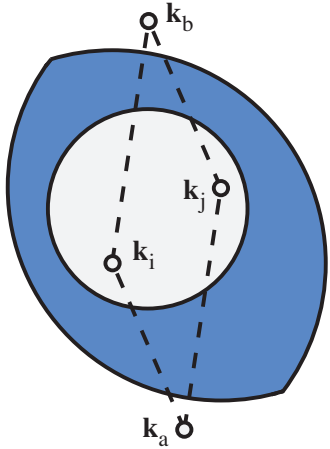}} \quad\quad\quad
\subfloat[This is an illustration of the \textit{union} $E_g$-cutoff. A second and different solution is to extend the allowed space of \reffig{fig:G1} to anywhere that either \{$\bfk_i+\bfg$\} or \{$\bfk_j+\bfg^\prime$\} would be allowed, also restoring the permutational symmetry]{\label{fig:G3}\includegraphics[width=0.4\textwidth]{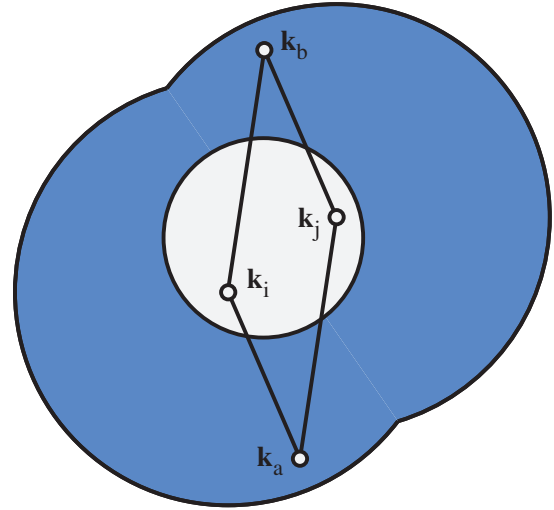}}
  
  \caption{Discussion and diagrams of cutoffs using momentum transfer vectors. The white circle represents the Fermi sphere, and the volume excluded from the virtual space by occupation effects is not considered.}
\end{figure*}

In this section, we develop a different type of basis set truncation for the HEG, based on the momentum transfer vector. Rather than the conventional definition of a single basis set for the whole calculation, we take the view that the basis set can be defined differently for each electron or each electron pair. This definition is not unique, even given a spherical cutoff, and we develop three types of basis set truncations showing that there is one that gives more rapid convergence to the CBS limit. This has the physical equivalence in reciprocal space of smearing out the rigid spherical cutoff into the surrounding space. This is motivated by a physical picture that electron coalescences should be treated on the same footing in momentum space.

\subsection{Introducing the momentum transfer vector}

Considering a general same-spin electron-electron-hole-hole excitation ${ij} \rightarrow {ab}$ connected by a matrix element $\bra \bfk_i \bfk_j | \potv | \bfk_a \bfk_b \ket- \bra \bfk_i \bfk_j | \potv | \bfk_b \bfk_a \ket$. We can therefore define two momentum transfer vectors for the excitation, $\bfg$ and $\bfg^\prime$\footnote{The momentum transfer vector is more commonly represented as {\bf q} in the literature},
\begin{equation}
\bfk_a=\bfk_i+\bfg \quad;\quad \bfk_b=\bfk_j-\bfg,
\end{equation}
\begin{equation}
\bfk_a=\bfk_j-\bfg^\prime \quad;\quad \bfk_b=\bfk_i+\bfg^\prime,
\end{equation}
where the allowed $\bfg$ vectors are such that $\bfk_a$ and $\bfk_b$ are both not in the occupied manifold. It is possible to re-write the sum over $\bfk_a$ and $\bfk_b$ in \refeq{K3CUT} in terms of these vectors,
\begin{equation}
E_\text{MP2} \left( k_c \right) = 
\sum_{\substack{0 \leq k_i \leq k_f \\ 0 \leq k_j \leq k_f  }}
\sum_{\substack{k_f < | \bfk_i+\bfg | \leq k_c \\ k_f < | \bfk_i+\bfg^\prime | \leq k_c }}
\delta_{\bfg,\bfk_j-\bfg^\prime-\bfk_i}  \frac{ \left( 2 ~ \vft{ \bfg  }  ^2 - \vft{ \bfg} \vft{\bfg^\prime} \right)  }{\Delta \epsilon_{ijab}}
\label{EMP2KC}
\end{equation}
where similar to before $\bfg^\prime$ is specified uniquely by $\bfg$, $\bfk_i$ and $\bfk_j$ using $\delta_{\bfg,\bfk_j-\bfg^\prime-\bfk_i}$.

By analogy with previous work in solid-state systems\cite{Marsman2009,Grueneis2010}, we now consider cutoffs that limit the extent of the momentum transfer vectors, and as such we impose a cutoff on the g-vectors such that they do not exceed a kinetic energy $E_g=\half g_c^2$, and such that $\bfk_a$ and $\bfk_b$ never reach the $E_k$-cutoff value $k_c$. The upper limit in the sum becomes entirely determined by $g_c$:
\begin{equation}
E_\text{MP2} \left( g_c \right) = 
\sum_{\substack{0 \leq k_i \leq k_f \\ 0 \leq k_j \leq k_f  }}
\sum_{\substack{k_f < | \bfk_i+\bfg |  \\ k_f < | \bfk_i+\bfg^\prime | \\ g \leq g_c \\ g^\prime \leq g_c}}
\delta_{\bfg,\bfk_j-\bfg^\prime-\bfk_i} \frac{ \left( 2 ~ \vft{ \bfg  }  ^2 - \vft{ \bfg} \vft{\bfg^\prime} \right)  }{\Delta \epsilon_{ijab}}.
\label{eq:Gpre}
\end{equation}
This gives us a new form of basis set truncation whose behaviour in the large $g$ limit might be different to $E_\text{MP2} \left( k_c \right)$ (\refeq{EMP2KC}), which we will now investigate.

It is also possible to remove the upper limits on the sums, replacing them in with radially symmetric step functions in k-space, 
\begin{equation}
\Theta \left(g-g_c\right) = \left\{
\begin{array}{ll}
1, & | \bfg | \leq g_c \\
0, & \text{otherwise}
\end{array}
\right. 
\end{equation}
yielding,
\begin{equation}
\begin{split}
&E_\text{MP2} = 
\sum_{\substack{0 \leq k_i \leq k_f \\ 0 \leq k_j \leq k_f  }}
\sum_{\substack{k_f < | \bfk_i+\bfg |  \\ k_f < | \bfk_i+\bfg^\prime |}}
\delta_{\bfg,\bfk_j-\bfg^\prime-\bfk_i} \frac{ 2 ~ \vft{ \bfg  }  ^2  \Theta \left(g-g_c\right)  }{\Delta \epsilon_{ijab}} \\
&-
\sum_{\substack{0 \leq k_i \leq k_f \\ 0 \leq k_j \leq k_f  }}
\sum_{\substack{k_f < | \bfk_i+\bfg |  \\ k_f < | \bfk_i+\bfg^\prime |}}
 \delta_{\bfg,\bfk_j-\bfg^\prime-\bfk_i}\frac{ \vft{ \bfg} \vft{\bfg^\prime} \Theta \left(g-g_c\right) \Theta \left(g^\prime-g_c\right)   }{\Delta \epsilon_{ijab}}
\label{eq:G1}
\end{split}
\end{equation}

This cutoff is shown diagrammatically in \reffig{fig:G1} for a specific electron pair, $\bfk_i$ and $\bfk_j$, illustrating that the basis set used to represent the virtual manifold is now no longer consistent between different electron pairs. By allowing \{$\bfg$\} and \{$\bfg^\prime$\} to span a certain range in reciprocal space, the virtual space represented by the sets \{$\bfk_i+\bfg$\} and \{$\bfk_j+\bfg^\prime$\} span a range dependent on $\bfk_i$ and $\bfk_j$ respectively. 

More severely than this, the basis set we have defined is also different for each electron. Considering a specific single same-spin excitation $\bra \bfk_i \bfk_j | \potv | \bfk_a \bfk_b \ket- \bra \bfk_i \bfk_j | \potv | \bfk_b \bfk_a \ket$ this means that sometimes the shorter momentum transfer vector is allowed while the longer is disallowed. This is shown diagrammatically in \reffig{fig:exG1}. 

%
Since this cutoff takes the view that each electron has its own basis set, this will be termed the \emph{local} $E_g$-cutoff. In \refeq{eq:G1} this is represented by the different ranges of the sums over the direct and exchange-like terms. In the exchange-like term the product of the step-functions serves to disallow some longer-momentum events. The implication of this is that in the general hole pair function space $| \bfk_a \bfk_b \ket$ can be allowed while its permutation $\hat{P}_{12}| \bfk_a \bfk_b \ket = - |\bfk_b \bfk_a \ket$ can be absent. This implies that not all terms accounted for in the direct term are properly balanced by the exchange-like term, and the antisymmetry of the wave function is ultimately not properly restored. 

In the conventional basis set scheme described in \refsec{sec1}, there is a variational principle: for a finite basis set you are guaranteed to not retrieve more correlation energy than the complete basis set limit. As the basis set is enlarged, the correlation energy is systematically lowered to the complete basis set limit correlation energy. This variationality is broken by use of a local $E_g$-cutoff. 

We can define two further ways of defining a $E_g$-cutoff, which do not suffer from these limitations. In the \emph{intersection} $E_g$-cutoff, we force the direct term to be removed from the sum if the exchange-like is rejected for the same \{$\bfg$,$\bfg^\prime$\} pair:

\begin{equation}
\begin{split}
&E_\text{MP2} = 
\sum_{\substack{0 \leq k_i \leq k_f \\ 0 \leq k_j \leq k_f  }}
\sum_{\substack{k_f < | \bfk_i+\bfg |  \\ k_f < | \bfk_i+\bfg^\prime |}}
\delta_{\bfg,\bfk_j-\bfg^\prime-\bfk_i} \frac{ 2 ~ \vft{ \bfg  }  ^2  P\left(\bfg,\bfg^\prime\right)  }{\Delta \epsilon_{ijab}} \\
&-
\sum_{\substack{0 \leq k_i \leq k_f \\ 0 \leq k_j \leq k_f  }}
\sum_{\substack{k_f < | \bfk_i+\bfg |  \\ k_f < | \bfk_i+\bfg^\prime |}}
\delta_{\bfg,\bfk_j-\bfg^\prime-\bfk_i}\frac{ \vft{ \bfg} \vft{\bfg^\prime}  P\left(\bfg,\bfg^\prime\right)  }{\Delta \epsilon_{ijab}}
\end{split}
\end{equation}
where $P\left(\bfg,\bfg^\prime\right)$ is given by,
\begin{equation}
P\left(\bfg,\bfg^\prime\right)=\Theta \left(g-g_c\right) \Theta \left(g^\prime-g_c\right).
\label{intersectionG}
\end{equation}
which can be thought of as a masking function in that it disallows certain electron pairs from being connected to different parts of the virtual space. 

In the \emph{union} $E_g$-cutoff, we force the exchange-like term to be preserved if the exchange-like term is rejected for the same \{$\bfg$,$\bfg^\prime$\} pair by use of,
\begin{equation}
P\left(\bfg,\bfg^\prime\right)=\Theta \left(g-g_c\right) +\Theta \left(g^\prime-g_c\right) -\Theta \left(g-g_c\right)\Theta \left(g^\prime-g_c\right),
\label{unionG}
\end{equation}
where the term $\Theta \left(g-g_c\right)\Theta \left(g^\prime-g_c\right) $ prevents double-counting when $\Theta \left(g-g_c\right)$ and $\Theta \left(g^\prime-g_c\right) $ are both 1.

The cutoffs are named after how they are generated from the sets \{$\bfk_i+\bfg$\} and \{$\bfk_j+\bfg^\prime$\}, shown in \reffig{fig:G2} and \reffig{fig:G3}.

\subsection{Comparison of the different cutoffs}

\begin{figure}[h]

\psfrag{KKK}[l][l][0.95][0]{$E_k$-cutoff} 
\psfrag{GGG1}[l][l][0.95][0]{Union $E_g$-cutoff} 
\psfrag{GGG2}[l][l][0.95][0]{Intersection $E_g$-cutoff} 
\psfrag{GGG3}[l][l][0.95][0]{Local $E_g$-cutoff} 
\psfrag{AAA}[l][l][0.95][0]{CBS $\pm$ 1\%} 
\psfrag{BBB}[l][l][0.95][0]{CBS result} 
\psfrag{XXX}[][][0.95][0]{$M^{-1}$} 
\psfrag{YYY}[][][0.95][0]{Correlation energy / a.u.} 
\includegraphics[width=0.45\textwidth]{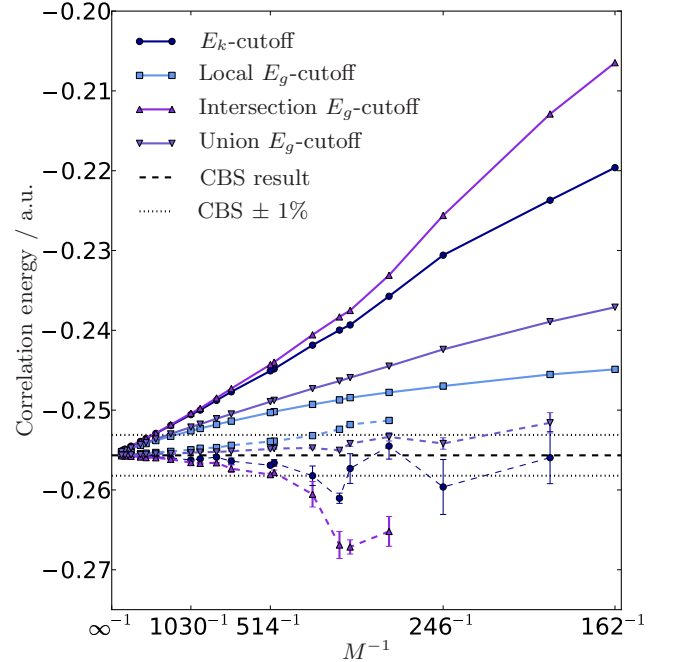}
\caption{Comparison of correlation energy retrieved as a function of basis set size for a variety of cutoff schemes.}
\label{loadsofGcutoffs}

\end{figure}

\begin{figure*}
\begin{centering}

\subfloat[]{

\psfrag{KKK1}[l][l][1.2][0]{$r_s=0.5$~a.u.} 
\psfrag{KKK2}[l][l][1.2][0]{$r_s=1.0$~a.u.} 
\psfrag{KKK3}[l][l][1.2][0]{$r_s=2.0$~a.u.} 
\psfrag{KKK4}[l][l][1.2][0]{$r_s=5.0$~a.u.} 
\psfrag{KKK5}[l][l][1.2][0]{$r_s=10.0$~a.u.} 
\psfrag{KKK6}[l][l][1.2][0]{$r_s=20.0$~a.u.} 

\psfrag{AAA}[l][l][1.2][0]{CBS $\pm$ 1\%} 
\psfrag{BBB}[l][l][1.2][0]{CBS result} 
\psfrag{XXX}[][][1.2][0]{$M^{-1}$} 
\psfrag{YYY}[][][1.2][0]{Fraction of correlation energy retrieved} 

\includegraphics[width=0.5\textwidth]{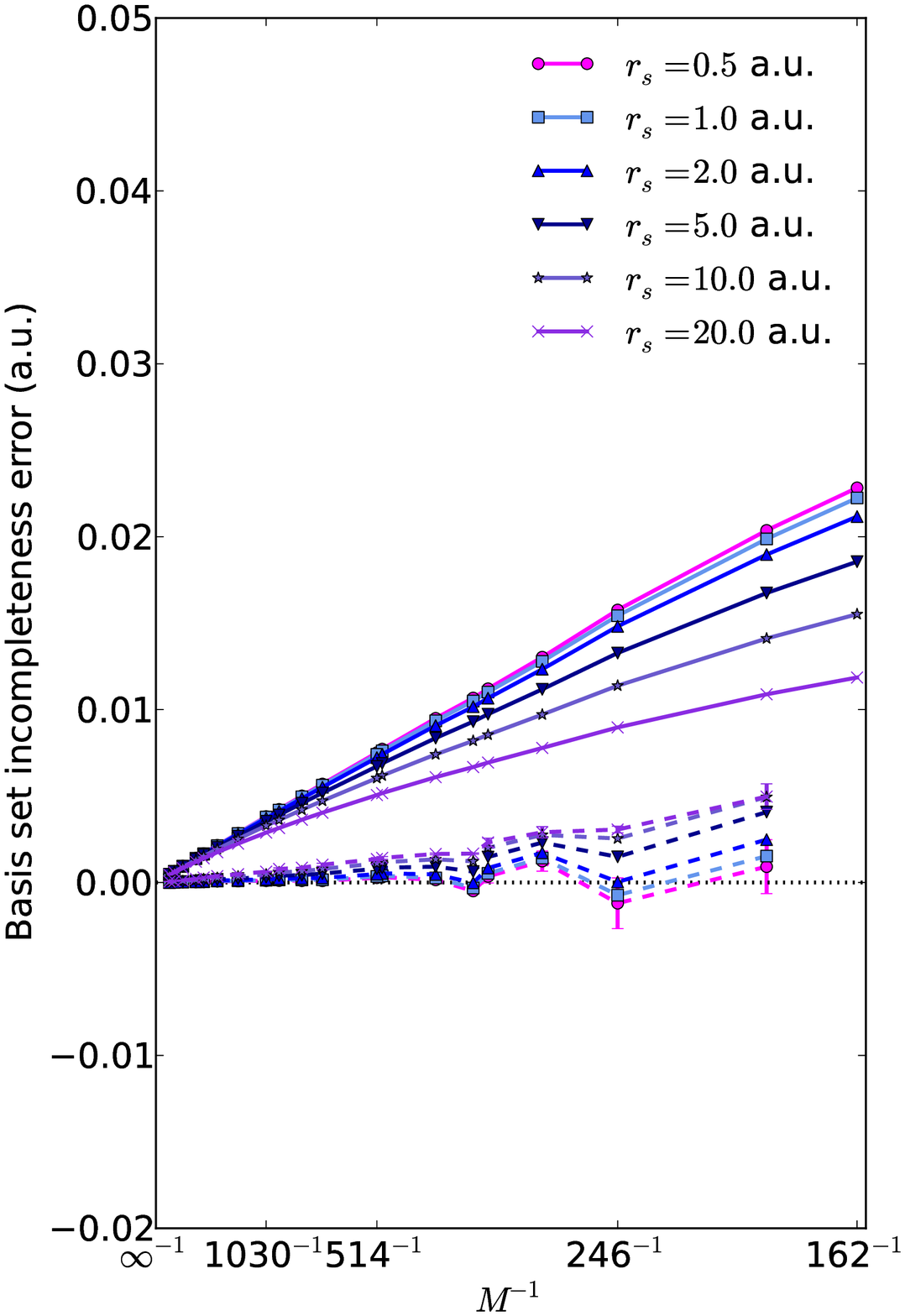}

}
\subfloat[]{

\psfrag{KKK1}[l][l][1.2][0]{$r_s=0.5$~a.u.} 
\psfrag{KKK2}[l][l][1.2][0]{$r_s=1.0$~a.u.} 
\psfrag{KKK3}[l][l][1.2][0]{$r_s=2.0$~a.u.} 
\psfrag{KKK4}[l][l][1.2][0]{$r_s=5.0$~a.u.} 
\psfrag{KKK5}[l][l][1.2][0]{$r_s=10.0$~a.u.} 
\psfrag{KKK6}[l][l][1.2][0]{$r_s=20.0$~a.u.} 

\psfrag{AAA}[l][l][1.2][0]{CBS $\pm$ 1\%} 
\psfrag{BBB}[l][l][1.2][0]{CBS result} 
\psfrag{XXX}[][][1.2][0]{$M^{-1}$} 
\psfrag{YYY}[][][1.2][0]{Fraction of correlation energy retrieved} 

\includegraphics[width=0.5\textwidth]{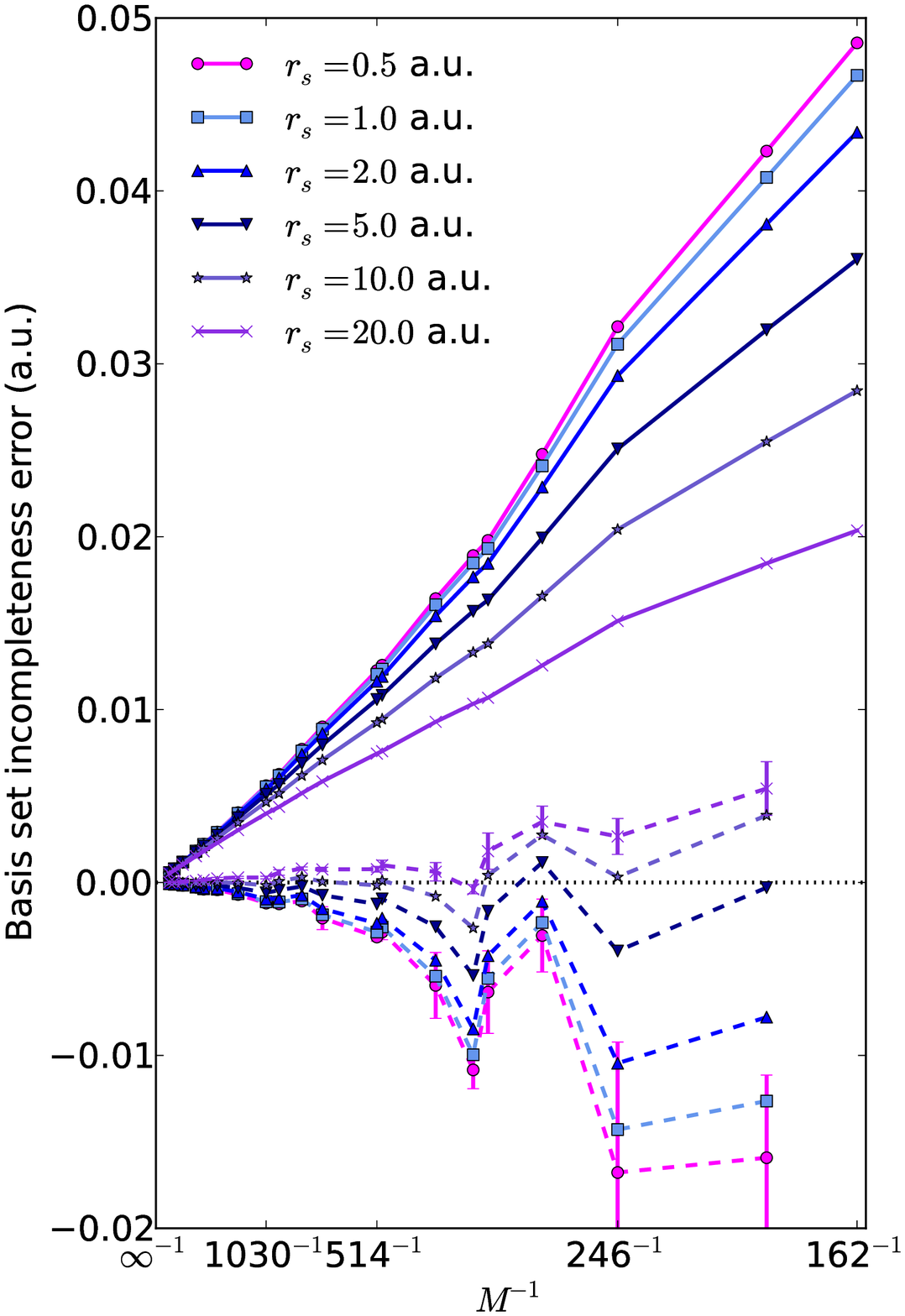}

}

\caption{Graphs comparing (a) union $E_g$-cutoff and (b) $E_k$-cutoff for a range of different densities. As the density is lowered, the extrapolations become more distant from the CBS limit even for MP2 theory, due to a rising contribution from exchange in the Hartree-Fock orbital energies that is not well-behaved with respect to $M$. Furthermore, more pronounced finite size effects are seen. The extrapolated results shown by the dotted lines are only represented with error bars in two cases ($r_s=0.5$~a.u., $r_s=20.0$~a.u.) for clarity. Complete basis set limit energies from which these basis set incompleteness errors are derived are tabulated in Table \ref{onlytable}.}
\label{rs_scan}

\end{centering}
\end{figure*}

Figure \ref{loadsofGcutoffs} shows correlation energies with these different cutoff schemes. In the case of the $E_g$-cutoffs, the $M$ is the number of spin orbital basis functions in a sphere with radius $g_c$ centred at the $\Gamma$-point. In some cases, in particular the union $E_g$-cutoff, the number of basis functions used for the calculation is higher, but  we believe that there is no better parameterization of the size of the basis set than this effective $M$.

All of the schemes regardless of the cutoff scheme converge to the CBS limit ultimately as $1/M$. As the size of the basis set goes to infinite extent, all $E_g$-cutoffs ultimately tend back towards the $E_k$ picture since the displacement of the occupied $k$-points from the $\Gamma$-point becomes negligible and the lines become identical. For the intersection and local $E_g$-cutoff schemes the curves, however, merge only at very large basis sets.

The positioning of the curves of each $E_g$ basis set can now be compared with that of the corresponding $E_k$ basis set. Both the intersection and union $E_g$-cutoffs can be thought of as lying in a larger $E_k$-cutoff basis set and are variational upper bounds of this larger basis set energy. In the intersection $E_g$-cutoffs scheme, terms are effectively removed from electron pairs that are of significant distance from the $\Gamma$-point. Furthermore, all excitations lie within $g_c$ of the $\Gamma$-point, meaning that this basis set produces a variational upper-bound to the $E_k$-cutoff basis set of the same size ($k_c=g_c$). In contrast, the union $E_g$-cutoff augments the basis set for those electron pairs that are not at the $\Gamma$-point by including basis functions that can have as high an energy as $\half\left(g_c^2+k_f^2\right)$. As such, this is now a variational upper bound of the $E_k$ basis set that completely encloses the radius $\left(g_c^2+k_f^2\right)^\half$.

In contrast, the local $E_g$-cutoff is neither variationally bounded by the complete basis set limit nor any $E_k$-cutoff basis set. In general, it can be considered that it has fewer exchange-like terms than the corresponding union basis set, and as such will produce a lower correlation energy than all of the basis sets with the same cutoffs. Since the exchange-like terms in the correlation energy are positive and partly neglected, the correlation energy becomes more negative than for the corresponding union basis set. Although this seems advantageous in the first instance, as it seems to retrieve a greater fraction of the CBS correlation energy, already in \reffig{loadsofGcutoffs} it can be seen that there is a tendency for this curve to arc at low basis sets, and could even have a maximum point in extreme cases.

Each cutoff has a separate behavior when a $1/M$ behaviour is used to extrapolate the result from a series of finite basis calculations. As noted previously, the $E_k$-cutoff basis set suffers from strong finite size effects, causing the extrapolation to behave jaggedly around the CBS result. This can be thought of being due to trying to recreate a spherical cutoff with a cubic grid. In common with this, the intersection $E_g$-cutoff has even stronger shell-filling effects, which are more pronounced because this basis set is trying to recreate the overlap between two spheres with this cubic grid, a shape with an even smaller volume to surface ratio. The local and union $E_g$-cutoffs have much smoother convergences with $1/M$ and their extrapolated results converge much more smoothly to the CBS limit. This could be because we have replaced spheres in $k$-space with more complex objects, and also are summing in more excitations for a given $M$.

In conclusion, the union $E_g$-cutoff seems to have the most desirable properties: variationality, correct symmetry of the wavefunction, smoothness and speed of convergence and extrapolation. When the density, as represented by $r_s$, is changed we might expect these relationships between the cutoffs to change. In \reffig{rs_scan}, we have considered the fraction of the CBS correlation energy obtained by basis sets at both higher and lower densities ($r_s=0.5-20.0$~a.u.) for the union $E_g$-cutoff and $E_k$-cutoff. As $r_s$ is raised, the basis set extrapolation becomes increasingly distant from the CBS result at smaller basis set sizes. This is due to the rise in the contribution from the exchange-like term in the MP2 energy, which has a less well-defined convergence with respect to the $M$ parameter that we are using. Furthermore, finite size effects become more visible. From these graphs, it is possible to see that for this system the union $E_g$-cutoff does continue to be the cutoff of choice for the reasons outlined above. For completeness, values of the complete basis set correlation energy for MP2 are presented in Table \ref{onlytable}.

\begin{table}[h]
\begin{tabular}{ | c | c |  }
 \hline
$\quad$$r_s$ (a.u.)$\quad$ & $\quad$Correlation energy (a.u.)$\quad$ \\
\hline
\hline
0.5 & -0.575442(1) \\
1.0 & -0.499338(2)\\
2.0 & -0.398948(2)\\
5.0 & -0.255664(4)\\
10.0  & -0.163951(6)\\
20.0 & -0.09749(1)\\
\hline
\end{tabular}
  \caption{Values of the complete basis set limit MP2 correlation energy obtained by extrapolation for $N=14$. The error estimate, shown in brackets refers to the random error in the last digit from extrapolation.}
  \label{onlytable}
\end{table}


\section{Generalisation to other single-reference quantum chemical methods}
\label{quantumchem}

In this section, we seek to generalise the discussion above to other single-reference quantum chemical
methods, in particular the coupled-cluster doubles (CCD) theory\cite{Bartlett2007} and the random phase approximation plus
second-order screened exchange (RPA+SOSEX)\cite{Freeman1977}. In these methods, the energy estimator depends on amplitudes, all of which vary when the basis set size is changed. This is in marked contrast with MP2 theory, in which only those basis functions \emph{added} when a basis set is enlarged acquire new contributions to the energy. We will discuss two possible strategies for extrapolating the energy to the CBS limit. We will show that a direct extrapolation based on calculations at different basis set sizes is one method for achieving the CBS limit, and that both CCD and RPA+SOSEX correlation energies behave as $1/M$ in common with the MP2 correlation energy. However, this suffers from the same slow convergence of the $E_k$-cutoff strategy outlined from MP2.

In an attempt to emulate the more effective extrapolation to the CBS limit provided by the momentum transfer vector cutoff schemes, we introduce a new approach to this problem, \emph{single-point extrapolation}, in which the contributions to the energy from a single calculation are re-grouped according to their arrangement in reciprocal space to form energy estimates from effective basis set sizes. These smaller effective basis set energies are then used to provide an extrapolation to the CBS limit. Although it will be demonstrated that this approach does provide more effective convergence, amplitude `relaxation' as the basis set size increases causes a problem with CCD at $r_s=5.0$~a.u. However, this crucially also allows for the adaptation of extrapolation to solid state systems, where direct extrapolation is not only slower to converge but also made difficult to achieve by the PAW approximation.

\subsection{Direct extrapolation of CCD and RPA+SOSEX}

\begin{figure*}

\subfloat[]{

\psfrag{KKK1}[l][l][1.2][0]{MP2} 
\psfrag{KKK2}[l][l][1.2][0]{CCD} 
\psfrag{KKK3}[l][l][1.2][0]{RPA+SOSEX} 
\psfrag{KKK4}[l][l][1.2][0]{\iFCIQMC} 
\psfrag{XXX}[][][1.0][0]{$M^{-1}$} 
\psfrag{YYY}[t][t][1.0][0]{Correlation energy (a.u.)} 
\psfrag{X6}[][][1.0][0]{$66^{-1}$} 
\psfrag{X5}[][][1.0][0]{$114^{-1}$} 
\psfrag{X4}[][][1.0][0]{$186^{-1}$} 
\psfrag{X3}[l][l][1.0][0]{$358^{-1}$} 
\psfrag{X2}[r][r][0.9][0]{$682^{-1}$} 
\psfrag{XLX}[][][1.0][0]{$|$} 
\psfrag{X2X}[][][1.0][0]{$\infty^{-1}$} 
\includegraphics[width=0.45\textwidth]{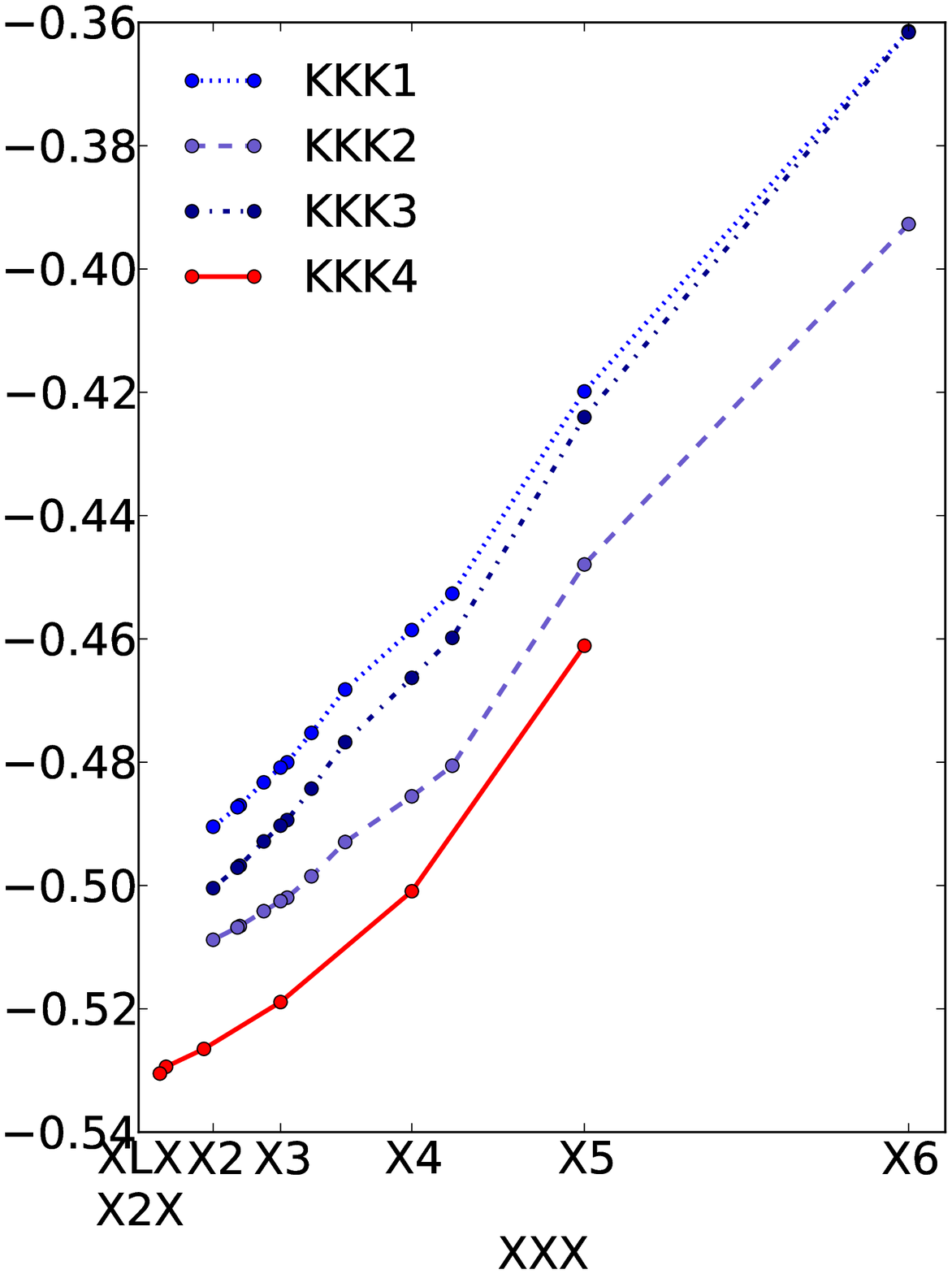}
} \quad\quad\quad
\subfloat[]{

\psfrag{KKK1}[l][l][1.2][0]{MP2} 
\psfrag{KKK2}[l][l][1.2][0]{CCD} 
\psfrag{KKK3}[l][l][1.2][0]{RPA+SOSEX} 
\psfrag{KKK4}[l][l][1.2][0]{\iFCIQMC} 
\psfrag{XXX}[][][1.0][0]{$M^{-1}$} 
\psfrag{YYY}[t][t][1.0][0]{Correlation energy (a.u.)} 
\psfrag{X6}[][][1.0][0]{$66^{-1}$} 
\psfrag{X5}[][][1.0][0]{$114^{-1}$} 
\psfrag{X4}[][][1.0][0]{$186^{-1}$} 
\psfrag{X3}[l][l][1.0][0]{$358^{-1}$} 
\psfrag{X2}[r][r][0.9][0]{$682^{-1}$} 
\psfrag{XLX}[][][1.0][0]{$|$} 
\psfrag{X2X}[][][1.0][0]{$\infty^{-1}$} 
\includegraphics[width=0.45\textwidth]{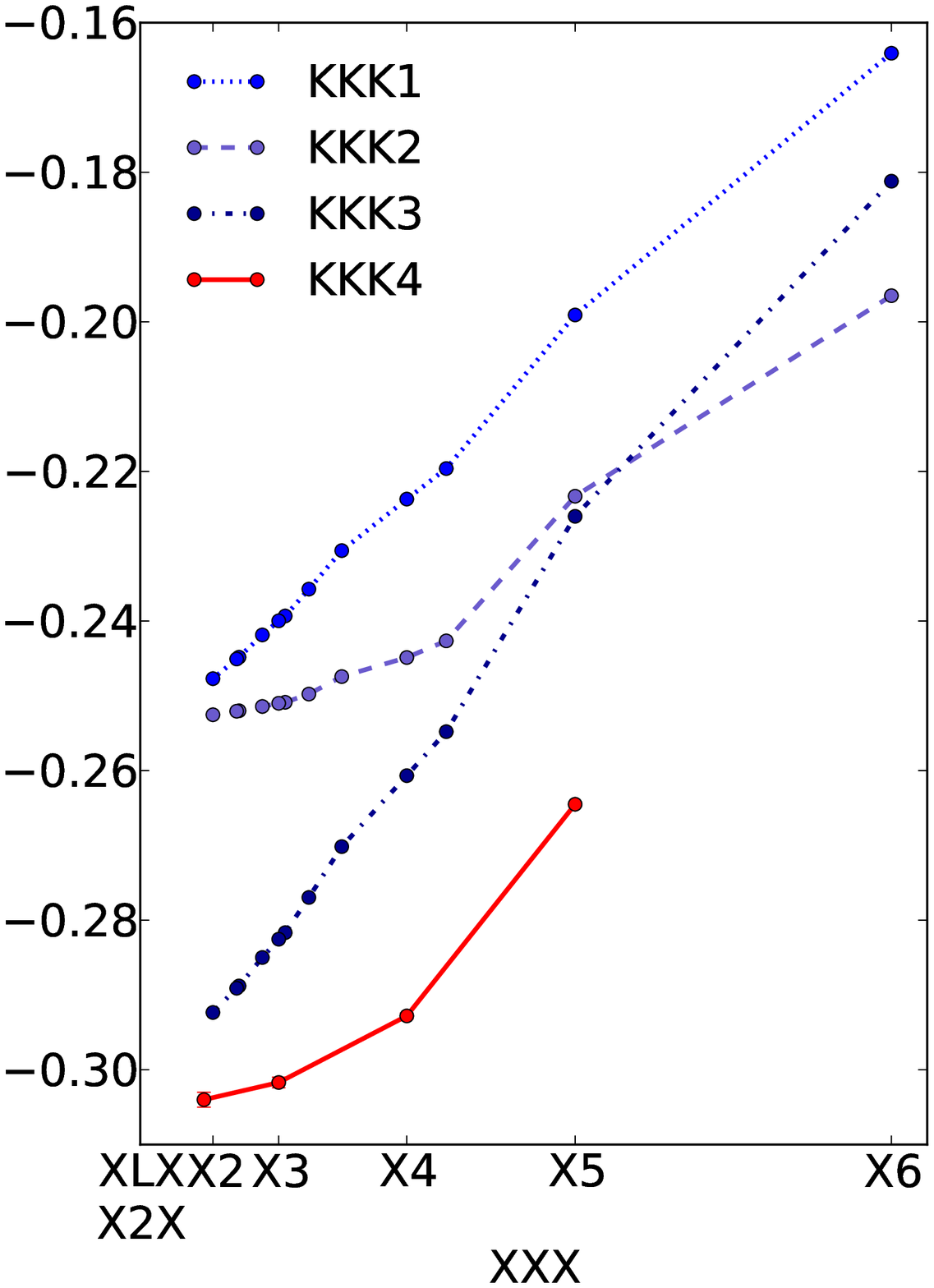}
\label{cross-over}

}

\caption{Comparison of correlation energy for the $N=14$ system for (a) $r_s=1.0$ (b) $r_s=5.0$ retrieved as a function of basis set size for a variety of quantum chemical methods.}\label{all_types}
\end{figure*}

In both CCD and RPA+SOSEX, the energy can be written in a configuration-space formalism as,
\begin{equation}
E_{\text{corr}} \left(M\right)=\sum_{ij}^{\text{occ}} \sum_{ab}^{M} \chi_{\bfk_i \bfk_j}^{\bfk_a \bfk_b} \left( M\right),
\end{equation}
where $\chi_{\bfk_i \bfk_j}^{\bfk_a \bfk_b}$ are the k-point labelled contributions to the energy which express a product of amplitudes on double excitations of the reference determinant and the corresponding Hamiltonian matrix element,
\begin{equation}
\chi_{\bfk_i \bfk_j}^{\bfk_a \bfk_b} = \left( 2 \vft{ \bfk_i-\bfk_a} - \vft{\bfk_j-\bfk_a} \right) t_{ij}^{ab}.
\label{eq:genproje}
\end{equation}
Appendix \ref{sec:ccd} contains a more detailled discussion of CCD and RPA+SOSEX and explains the evaluation of the respective $t_{ij}^{ab}$ amplitudes. The additional complexity compared to MP2 is that all $t_{ij}^{ab}$ vary when the basis set size is increased. Again we quantify the size of the basis set of the virtual orbitals using $M$. $M$ corresponds the total number of orbitals inside the cutoff sphere ($E_k$-cutoff in MP2). 
%

Performing calculations at different cutoffs allows direct extrapolation of the energy, with this behaving as $1/M$ in the large $M$ limit. Comparison between the finite basis-set energies retrieved by different quantum chemical methods is shown in \reffig{all_types} for the 14-electron problem at $r_s=1.0$ and $r_s=5.0$. All methods considered show a $1/M$ relationship in the high $M$ regime. In this graph, exact benchmarks from a new electronic structure method called initiator full configuration interaction quantum Monte Carlo (\iFCIQMCbracket) are presented for comparison from Ref.~\onlinecite{UEGPaper1}. This method utilises a stochastic algorithm to calculate FCI accuracy energies at greatly reduced computational cost\cite{FCIQMCPaper1,FCIQMCPaper2}. The gradient and onset of the $1/M$ behaviour varies with $r_s$ and method. At the higher $r_s$-value CCD best resembles the FCI behaviour, with MP2 and RPA+SOSEX resembling one another. All of the methods behave similarly with $M$ at the lower $r_s$-value. The ability of RPA+SOSEX to retrieve most of the FCI correlation energy at $r_s=5.0$ can be attributed in part to capturing too much (dynamic) correlation energy at high $M$. The cross-over between CCD and RPA+SOSEX (\reffig{cross-over}) highlights the difficulties of comparing methods at a finite basis set size. Clearly RPA+SOSEX or MP2 can not be used to estimate the finite size and basis set incompleteness error of \iFCIQMC or CCD, whereas CCD may be well suited to correct these errors in \iFCIQMCbracket. No attempt has been made to extrapolate these methods to the thermodynamic limit, in which MP2 is well-known to diverge, since this is beyond the scope of this paper. 
 
\subsection{Single-point extrapolation of CCD and RPA+SOSEX}

We now seek a momentum transfer vector cutoff scheme for CCD and RPA+SOSEX, in particular aiming to re-produce the properties of the union $E_g$-cutoff explored in \refsec{MTVCS}. After performing a single calculation in a basis set, 
\begin{equation}
E_{\text{corr}^\prime} \left(M \right)=\sum_{ij}^{\text{occ}} \sum_{ab}^{M^{\prime}} \chi_{\bfk_i \bfk_j}^{\bfk_a \bfk_b} \left(M \right).
\label{ecorr_kspace}
\end{equation}
Applying the masking function $P\left(\bfg,\bfg^\prime\right)$ defined previously for the union $E_g$-cutoff,
\begin{equation}
\begin{split}
P\left(\bfg,\bfg^\prime ; M^\prime  \right)&=\Theta \left(g-g_c\right) +\Theta \left(g^\prime-g_c\right) \\
&-\Theta \left(g-g_c\right)\Theta \left(g^\prime-g_c\right),
\end{split}
\end{equation}
which is associated with a new basis set size $M^\prime$  (described in \refsec{MTVCS}), to \refeq{ecorr_kspace} yields,
\begin{equation}
\begin{split}
E_{\text{corr},\text{eff}} \left(M, M^\prime \right)&=\sum_{ij}^{\text{occ}} \sum_{ab}^{M} \chi_{\bfk_i \bfk_j}^{\bfk_a \bfk_b} \left(M \right) \\
&\times P_g\left(\bfk_i-\bfk_a,\bfk_j-\bfk_a ; M^\prime \right),
\label{rebinning_equation}
\end{split}
\end{equation}
where we have explicitly noted that this formulation of the correlation energy is dependent on both $M^\prime$ and $M$. These correlation energies are labelled both by a true basis set size $M$ and what we will call an effective basis set size $M^{\prime}$. We now follow the procedure of performing a single calculation with $M$ spin orbitals, take the amplitudes and apply the relationship given in \refeq{rebinning_equation} for different values of $M^{\prime}$. 

Analyzing \refeq{rebinning_equation}, it is possible to see that there are two limiting values for $E_{\text{corr},\text{eff}}$. When $M^{\prime}=0$, the effective basis set correlation energy is zero, and when $M^{\prime}$ is such that all possible momentum transfers are included in the sum (when $g_c > k_c + k_f$), the effective basis set correlation energy is simply the basis set correlation energy $E_{\text{corr}}^\prime$ (\refeq{ecorr_kspace}).

In between these limits, if the amplitudes $t_{ij}^{ab}$ are always the opposite sign to the matrix element $\left( 2 \vft{ \bfk_i-\bfk_a  } -\vft{\bfk_j-\bfk_a} \right)$, there will be a monotonic decrease of $E_{\text{corr},\text{eff}} \left(M, M^{\prime} \right)$ to the basis set correlation energy as $M^{\prime}$ is increased. 

In MP2 theory, 
this monotonic decrease will be strictly observed, and can be shown to be identical to the union $E_g$-cutoff scheme when $k_c > g_c +k_f $. For this region, $0 < g_c  < k_c -k_f$, therefore the same tendency to follow a $1/M$ behavior will be seen. When $g_c  > k_c -k_f $, deviation from this behavior will be seen due to momentum transfer vectors being disallowed from not being in the original $k_c$ basis.

Unlike the previous formulation, this can now be applied to any method with an estimator of the form \refeq{eq:genproje}. However, since the amplitudes also depend on $M$, this is an approximation and convergence with this second cutoff should also be obtained.

\begin{figure}

\psfrag{KKK1}[l][l][1.0][0]{Effective energies} 
\psfrag{KKK2}[l][l][1.0][0]{Direct extrapolation} 
\psfrag{KKK3}[l][l][1.0][0]{Discarded points} 

\psfrag{AAA}[l][l][1.0][0]{CBS $\pm$ 1\%} 
\psfrag{BBB}[l][l][1.0][0]{CBS result} 
\psfrag{XXX}[][][1.0][0]{$M^{-1}$ or $M^{\prime -1}$} 
\psfrag{YYY}[][][1.0][0]{Correlation energy (a.u.)} 
\psfrag{X4}[][][1.0][0]{$54^{-1}$} 
\psfrag{X3}[l][l][1.0][0]{$294^{-1}$} 
\psfrag{X2}[][][1.0][0]{$682^{-1}$} 
\psfrag{XL}[][][1.0][0]{$|$} 
\psfrag{XX}[][][1.0][0]{$\infty^{-1}$} 

\includegraphics[width=0.45\textwidth]{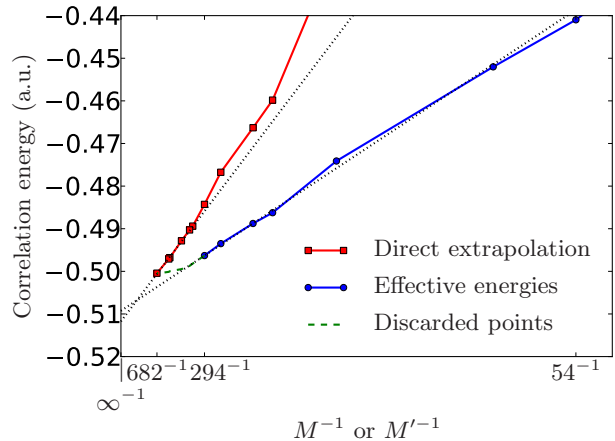}

\caption{Comparison between direct extrapolation and single-point extrapolation (SPE) for RPA+SOSEX on the $N=14$, $r_s=1.0$~a.u. gas. In the conventional direct extrapolation, calculations are performed at a series of basis set sizes $M$ and then extrapolated using a $1/M$ fit to the high $M$ limit. In the SPE, a single calculation is performed at an overall basis set size of $M$, in this case $M=682$, and effective basis set energies are constructed according to \refeq{rebinning_equation} over the full range of $M^{\prime}$. Some of these points are discarded as $M^{\prime}$ approaches $M$ since not all momentum transfer vectors can be accommodated within the basis set (dashed green line, discussed in the text). The extrapolations are shown by dotted lines, and agree in the CBS limit within reasonable extrapolation error estimates ($\sim 2\times 10^{-3}$a.u.).}
\label{general_SPE}
\end{figure}

\begin{figure*}
\begin{centering}

\subfloat[]{

\psfrag{KKK1}[l][l][1.0][0]{$E_\text{RPA+SOSEX}$} 
\psfrag{KKK2}[l][l][1.0][0]{Direct extrapolation} 
\psfrag{KKK3}[l][l][1.0][0]{$E_g$ SPE} 

\psfrag{AAA}[l][l][1.0][0]{CBS $\pm$ 1\%} 
\psfrag{BBB}[l][l][1.0][0]{CBS result} 
\psfrag{XXX}[][][1.0][0]{$M^{-1}$ or $M^{\prime -1}$} 
\psfrag{YYY}[][][1.0][0]{Correlation energy (a.u.)} 
\psfrag{X5}[][][1.0][0]{$114^{-1}$} 
\psfrag{X4}[][][1.0][0]{$186^{-1}$} 
\psfrag{X3}[l][l][1.0][0]{$358^{-1}$} 
\psfrag{X2}[][][1.0][0]{$682^{-1}$} 
\psfrag{XL}[][][1.0][0]{$\infty^{-1}$} 

\includegraphics[width=0.4\textwidth]{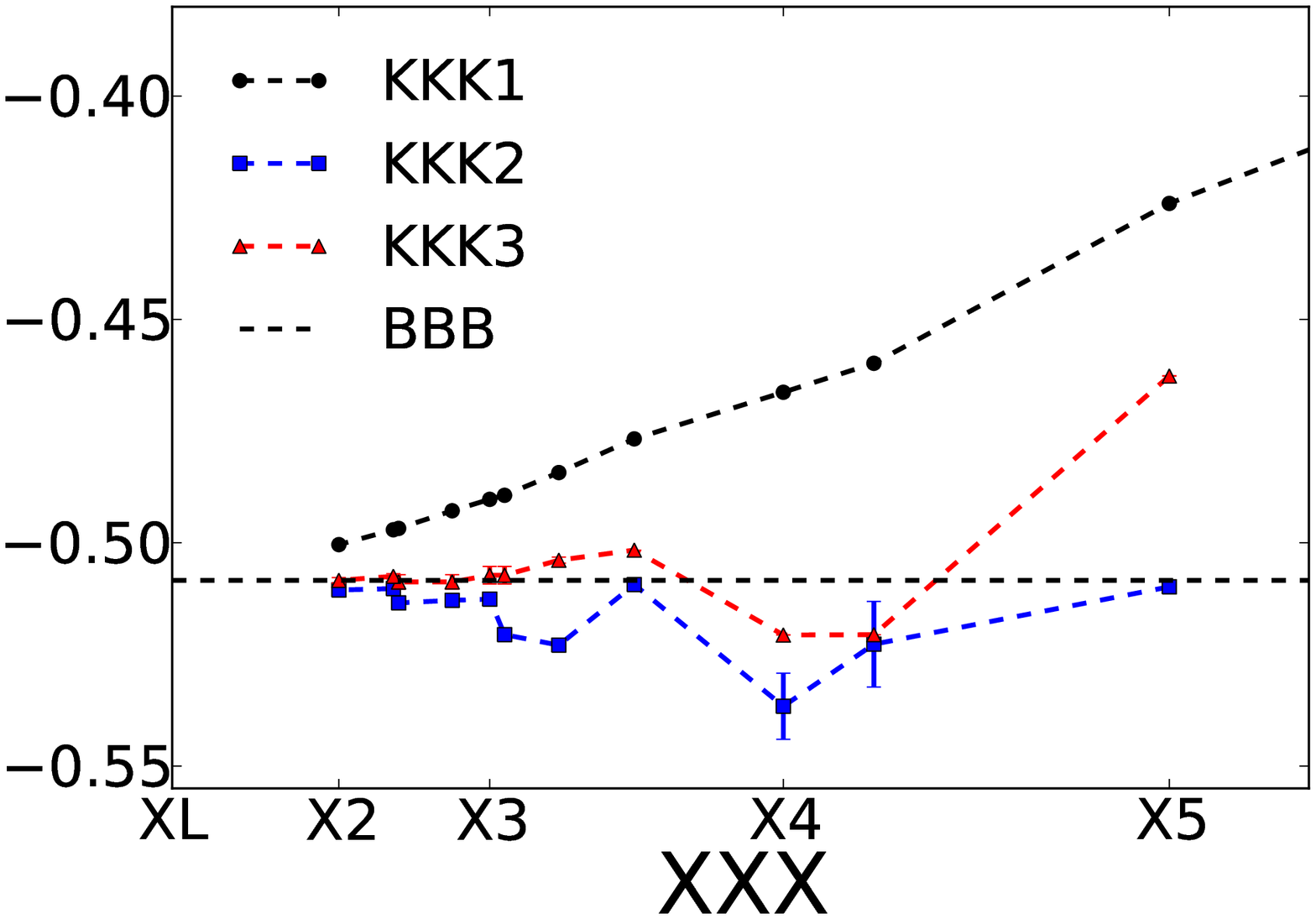}

}\quad\quad\quad
\subfloat[]{

\psfrag{KKK1}[l][l][1.0][0]{$E_\text{RPA+SOSEX}$} 
\psfrag{KKK2}[l][l][1.0][0]{Direct extrapolation} 
\psfrag{KKK3}[l][l][1.0][0]{$E_g$ SPE} 

\psfrag{AAA}[l][l][1.0][0]{CBS $\pm$ 1\%} 
\psfrag{BBB}[l][l][1.0][0]{CBS result} 
\psfrag{XXX}[][][1.0][0]{$M^{-1}$ or $M^{\prime -1}$} 
\psfrag{YYY}[][][1.0][0]{Correlation energy (a.u.)} 
\psfrag{X5}[][][1.0][0]{$114^{-1}$} 
\psfrag{X4}[][][1.0][0]{$186^{-1}$} 
\psfrag{X3}[l][l][1.0][0]{$358^{-1}$} 
\psfrag{X2}[][][1.0][0]{$682^{-1}$} 
\psfrag{XL}[][][1.0][0]{$\infty^{-1}$} 
\includegraphics[width=0.4\textwidth]{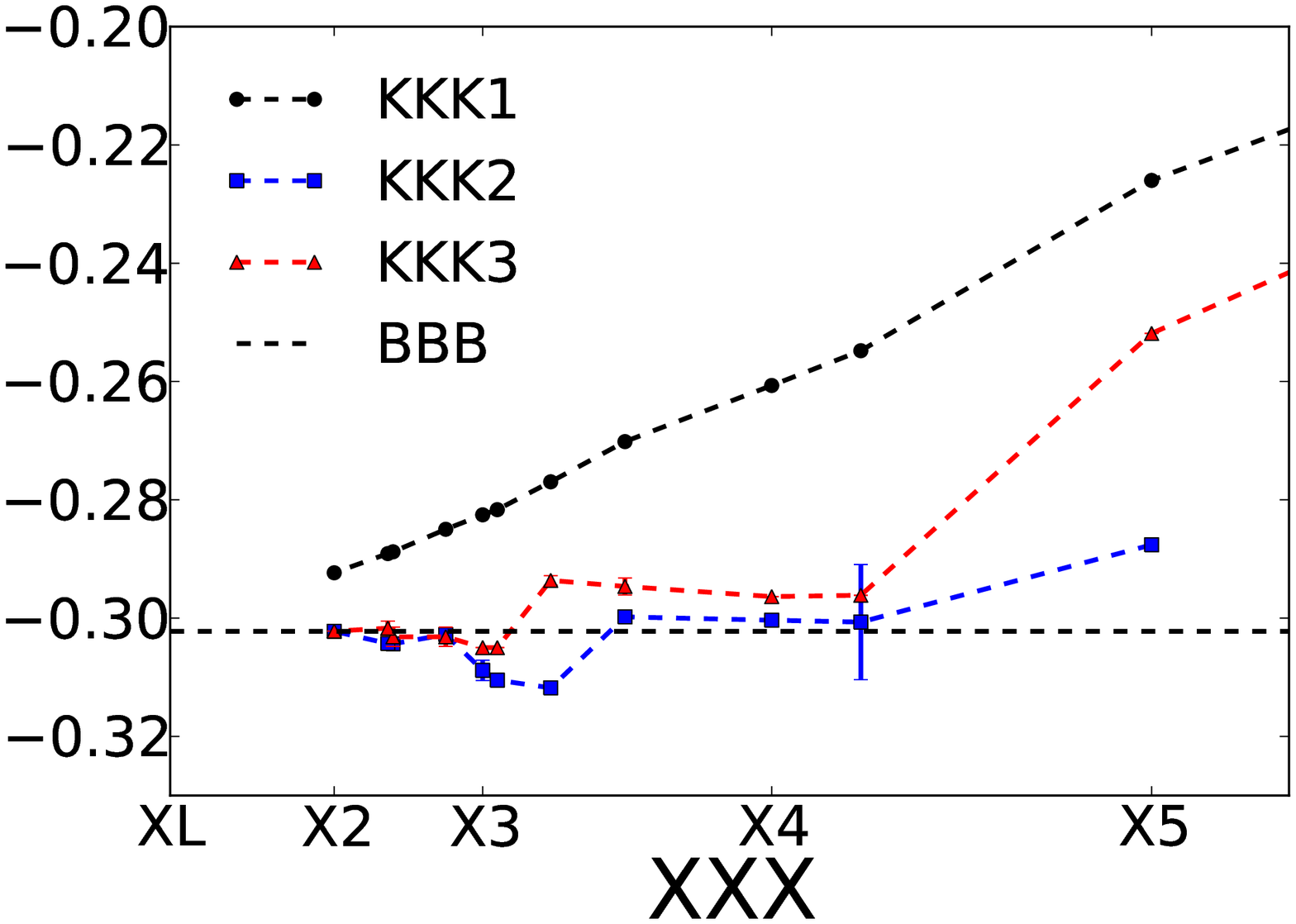}
}

\subfloat[]{

\psfrag{KKK1}[l][l][1.0][0]{$E_\text{CCD}$} 
\psfrag{KKK2}[l][l][1.0][0]{Direct extrapolation} 
\psfrag{KKK3}[l][l][1.0][0]{$E_g$ SPE} 

\psfrag{AAA}[l][l][1.0][0]{CBS $\pm$ 1\%} 
\psfrag{BBB}[l][l][1.0][0]{CBS result} 
\psfrag{XXX}[][][1.0][0]{$M^{-1}$ or $M^{\prime -1}$} 
\psfrag{YYY}[][][1.0][0]{Correlation energy (a.u.)} 
\psfrag{X5}[][][1.0][0]{$114^{-1}$} 
\psfrag{X4}[][][1.0][0]{$186^{-1}$} 
\psfrag{X3}[l][l][1.0][0]{$358^{-1}$} 
\psfrag{X2}[][][1.0][0]{$682^{-1}$} 
\psfrag{XL}[][][1.0][0]{$\infty^{-1}$} 

\includegraphics[width=0.4\textwidth]{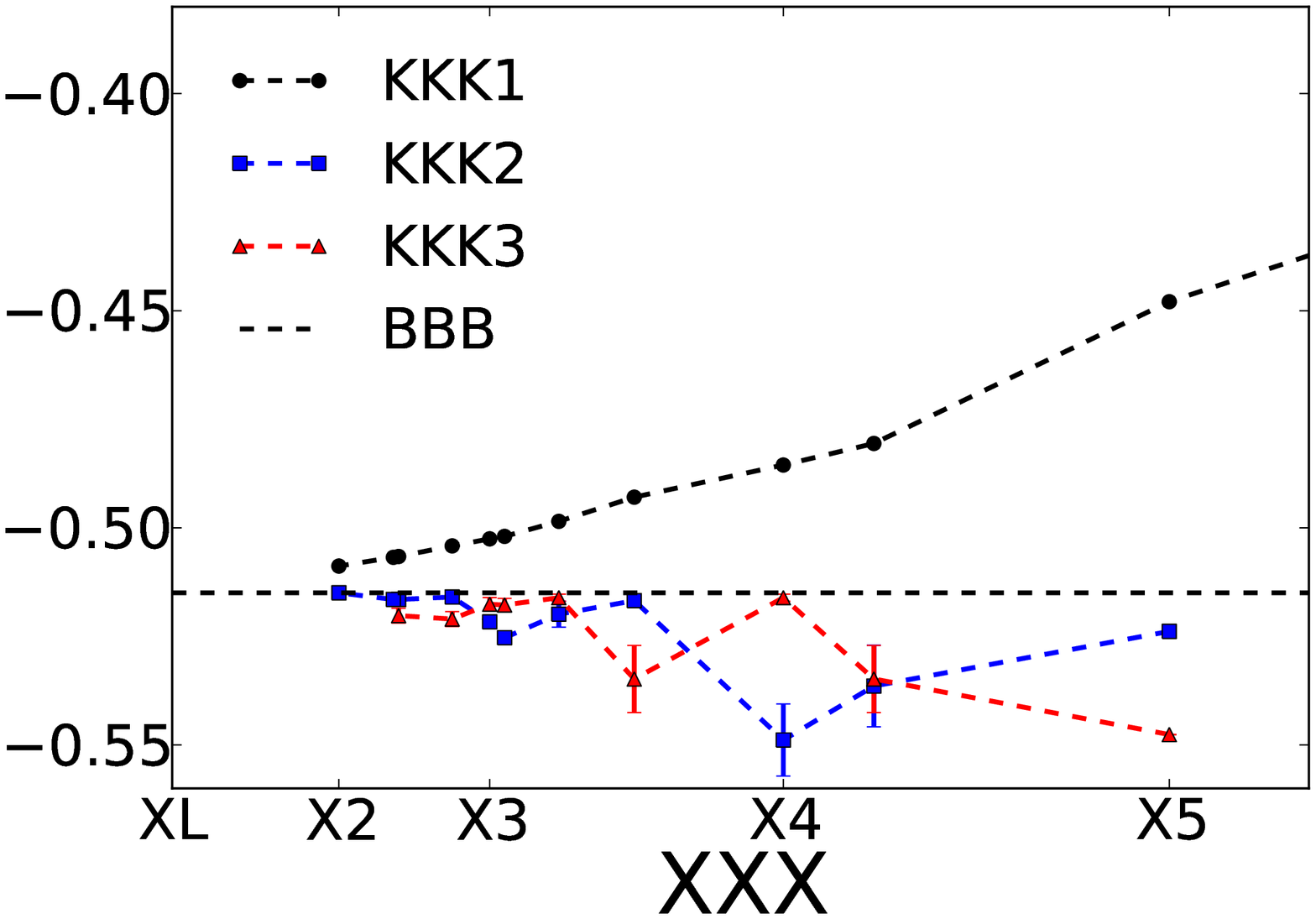}
}\quad\quad\quad
\subfloat[]{

\psfrag{KKK1}[l][l][1.0][0]{$E_\text{CCD}$} 
\psfrag{KKK2}[l][l][1.0][0]{Direct extrapolation} 
\psfrag{KKK3}[l][l][1.0][0]{$E_g$ SPE} 

\psfrag{AAA}[l][l][1.0][0]{CBS $\pm$ 1\%} 
\psfrag{BBB}[l][l][1.0][0]{CBS result} 
\psfrag{XXX}[][][1.0][0]{$M^{-1}$ or $M^{\prime -1}$} 
\psfrag{YYY}[][][1.0][0]{Correlation energy (a.u.)} 
\psfrag{X5}[][][1.0][0]{$114^{-1}$} 
\psfrag{X4}[][][1.0][0]{$186^{-1}$} 
\psfrag{X3}[l][l][1.0][0]{$358^{-1}$} 
\psfrag{X2}[][][1.0][0]{$682^{-1}$} 
\psfrag{XL}[][][1.0][0]{$\infty^{-1}$} 

\includegraphics[width=0.4\textwidth]{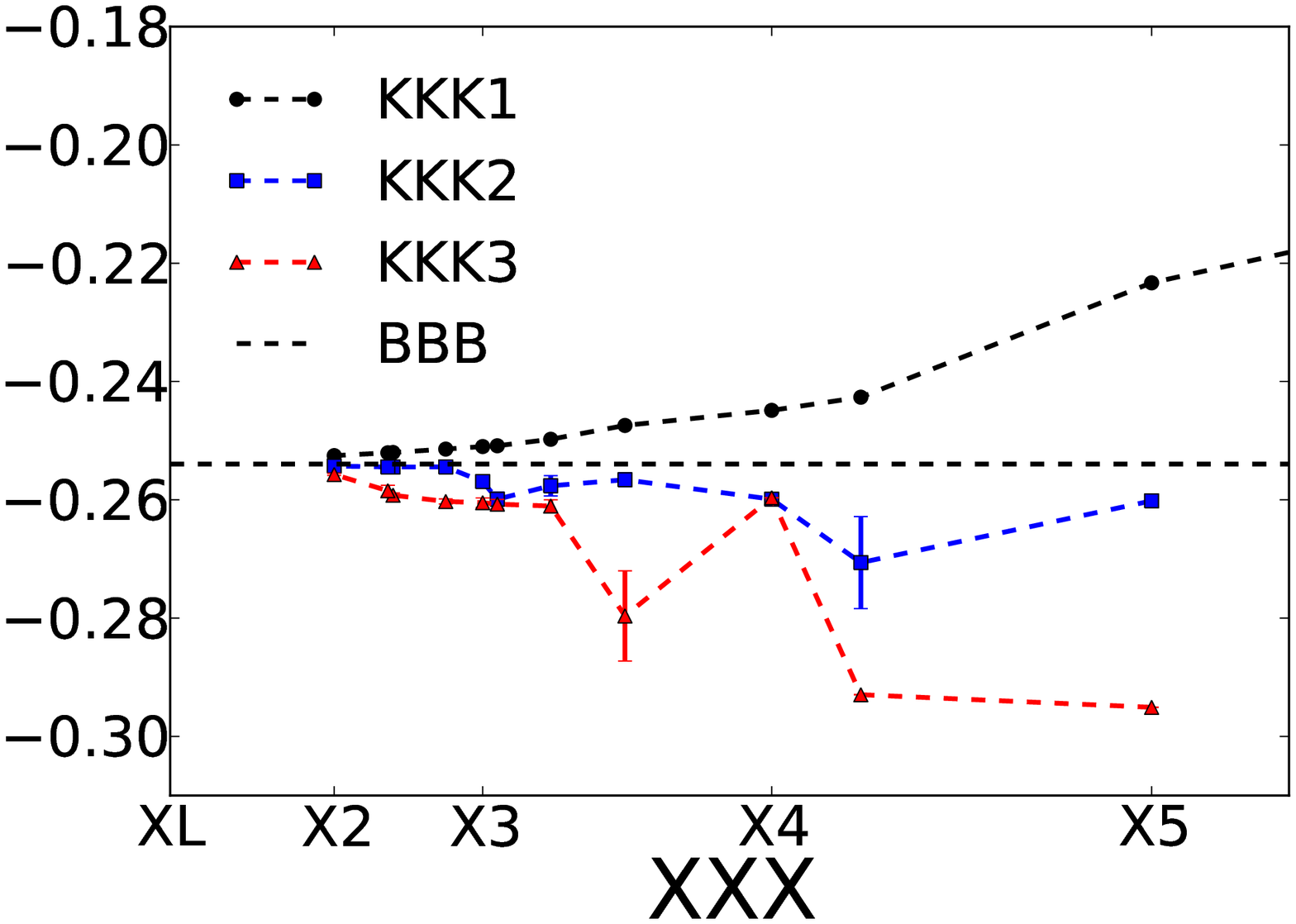}
}

\caption{Correlation energies calculated for the 14-electron system using RPA+SOSEX at (a) $r_s=1.0$ and (b) $r_s=5.0$ and using CCD (c) $r_s=1.0$ and (d) $r_s=5.0$. Direct extrapolation and single-point extrapolation (SPE) are compared at a variety of basis set sizes (SPE curves just show extrapolated results). SPE performs best for RPA at $r_s=1.0$, where it converges faster than direct extrapolation, and worst for CCD at $r_s=5.0$, where it converges slower than direct extrapolation.}
\label{RPA_CCD}

\end{centering}
\end{figure*}

Figure \ref{general_SPE} shows that these effective basis set energies have the property that they also converge as $1/M^{\prime}$ and can be used to extrapolate for a CBS estimate. \reffig{RPA_CCD} shows these extrapolations for RPA+SOSEX and CCD, comparing them with conventional direct extrapolation. In general, the RPA+SOSEX correlation energy converges faster using the $E_g$-cutoff single-point extrapolation than the $E_k$-cutoff direct extrapolation, which also has the advantage that only one calculation needs to be performed at a single basis set size. For CCD, this advantage is greatly obscured by finite size effects (which would become less for larger system sizes) and is not seen at all for $r_s=5.0$~a.u. due to flattening off of the finite basis set correlation energies and greater coefficient relaxation effects arising from stronger correlation.

Extensive discussion and analysis of relaxation effects are beyond the scope of this paper, but this method has also been successfully applied to the stochastic quantum chemical method \iFCIQMCbracket, and the further benefits of applying such a technique in a stochastic framework are discussed in Ref.~\onlinecite{UEGPaper2preprint}.

\section{Application to general solid state systems}

In this final methodological section, we discuss extrapolation schemes available to solid state calculations using a plane wave basis set.
We start by noting that for solid state systems the previous methodology of an $E_k$-cutoff (\refsec{sec1}), or
equivalently, an $M$ `true' basis set (\refsec{quantumchem}) is not easily defined.
Previous work to resolve this in a plane-wave basis set has used a resolution of the identity basis set to identify
Hamiltonian matrix elements. Following a similar argument made in this paper, previous authors have found that
the correlation energy converges with respect to this auxiliary basis set as $1/M$.
However, this greatly resembles the \emph{local} $E_g$-cutoff described
in this \refsec{MTVCS} and with the most severe penalty being that it is not variational with the CBS limit.\cite{Marsman2009}
We therefore examine
the improved extrapolation strategies based on the single-point extrapolation scheme discussed previously (union $E_g$-cutoff and $E_k$-cutoff),
which we believe to restore variationality and correct symmetry properties. Finally, we show how this can be applied to an example periodic system.

\subsection{Formulation of the single point extrapolation for solid state systems}

The correlation energy expression in general wavefunction based methods is given by
\begin{equation}
E_{\rm corr}\left(M\right)=\sum_{ij} \sum_{ab}^{M} t_{ij}^{ab} (2 v_{ij}^{ab} - v_{ij}^{ba})^*.
\label{eq:corren}
\end{equation}
The indices $i,j,k$ and $a,b,c,d$ refer to occupied and unoccupied orbitals, respectively
and are understood to be a shorthand for the band index and Bloch wavevector.
In contrast to the homogeneous electron gas, however, the orbitals are no longer constituted by plane waves and
correspond to eigenfunctions of the respective Hartree--Fock (HF) or Kohn--Sham (KS) one-electron Hamiltonians.
$M$ corresponds to the number of basis functions used in the description of occupied as well as unoccupied orbitals.
$v_{ij}^{ab}$ and $t_{ij}^{ab}$ refer to electron repulsion integrals and many-electron wavefunction amplitudes, respectively.
\begin{equation}
v_{ij}^{ab}=e^2\int \frac{
\langle\psi_i|{\bf r}\rangle\langle{\bf r}|\psi_a\rangle
\langle\psi_j|{\bf r}'\rangle\langle{\bf r}'|\psi_b\rangle}
{|{\bf r}-{\bf r}'|} {\rm d}{\bf r}'{\rm d}{\bf r}.
\end{equation}
For the sake of brevity, we have neglected single excitation (SE) contributions to the correlation energy in Eq.~(\ref{eq:corren}).
Depending on the approximation and reference determinant used in calculating the wavefunction, SE contributions might have to be included
but do not modify any of the conclusions drawn below.

We now seek to apply the previously outlined \emph{union} $E_g$ and $E_k$ single-point extrapolation scheme to general solid state systems.
To this end we introduce a projection matrix that transforms the HF/KS-orbitals onto a plane-wave basis set and reads
\begin{equation}
U_{n\bf G}=\langle \phi_n | \bf G \rangle,
\label{eq:defU}
\end{equation}
where $|\bf G \ket$ is a plane wave, $e^{i\bf G r}$, and $\phi_n$ constitutes a HF/KS orbital.
If no $k$-point sampling is used, {\bf G} corresponds to a reciprocal lattice vector that lies within a given spherical cutoff. For arbitrary $k$-point meshes, {\bf G} refers to a linear combination of a reciprocal lattice vector and the Bloch wavevector of the corresponding orbital $\phi_n$. As such, the following equations can all be implemented in the framework of a fully periodic code that samples arbitrary $k$-point meshes straight forwardly. In this study we will, however, restrict ourselves to $\Gamma$-point only calculations.
We note that
\begin{equation}
\delta_{nm}=\sum_{\bf G} U_{n\bf G} U^{-1}_{{\bf G} m}.
\label{eq:ident}
\end{equation}
If the employed finite plane-wave basis set is complete and large enough to span the space of
all orbitals $\phi_n$, $U_{n\bf G}$ becomes a unitary matrix. However, in our case, we use fewer orbitals than plane waves. We typically choose a plane-wave basis set for which $U_{n\bf G}$ is not full rank, and calculate $U^{-1}_{n\bf G}$ using a singular value decomposition.

Inserting Eq.~(\ref{eq:ident}) into Eq.~(\ref{eq:corren}) gives
\begin{equation}
E_{\rm corr}=\sum_{i} \sum_{{\bf G},{\bf G'},{\bf G''}} \tilde{t}_{i\bf G''}^{\bf G G'} (2 \tilde{v}_{i\bf G''}^{\bf G G'} - \tilde{v}_{i\bf G''}^{\bf G' G}),
\label{eq:ecplanew}
\end{equation}
where
\begin{equation}
\tilde{v}_{i\bf G''}^{\bf G G'}=\sum_{jab} {U^{-1}}^*_{{\bf G''} j} U^{-1}_{{\bf G} a} U^{-1}_{{\bf G'} b}v_{ij}^{ab}
\end{equation}
and
\begin{equation}
\tilde{t}_{i\bf G''}^{\bf G G'}=\sum_{jab} {U}^*_{j\bf G''} U_{a\bf G} U_{b\bf G'}t_{ij}^{ab}.
\end{equation}
In contrast to Eq.~(\ref{eq:corren}), Eq.~(\ref{eq:ecplanew}) is suitable for the extrapolation schemes described in Sec.~\ref{MTVCS},
since the indices ${\bf G, G'}$ and ${\bf G ''}$ refer again to plane-waves. Inserting a masking function that
has been introduced for the union $E_g$ cutoff into Eq.~(\ref{eq:ecplanew}) gives
\begin{equation}
\label{eq:ecplanew2}
\begin{split}
E_{\rm corr,eff}(M,M^{\prime})&=\sum_{i} \sum_{{\bf G},{\bf G'},{\bf G''}} \chi_{i\bf G''}^{\bf G G'}(M)
\\ &\quad\times P_g({\bf G'}-{\bf G''},{\bf G}-{\bf G''};M^{\prime}),
\end{split}
\end{equation}
where
\begin{equation}
\chi_{i\bf G''}^{\bf G G'}(M)=\tilde{t}_{i\bf G''}^{\bf G G'} (2 \tilde{v}_{i\bf G''}^{\bf G G'} - \tilde{v}_{i\bf G''}^{\bf G' G}).
\end{equation}
Note that only three out of four orbital indices are transformed, and that the transformed $\chi$ is not symmetric.
Due to momentum conservation in the transformed basis, the (truncated) correlation energies obtained are, however,
invariant with respect to the transformation of $i$. Note that $E_{\rm corr,eff}(M,M^{\prime})$ converges towards Eq.~(\ref{eq:corren}) for a sufficiently large $M^{\prime}$. Replacing $P_g({\bf G'}-{\bf G''},{\bf G}-{\bf G''};M^{\prime})$ with $P_k({\bf G'},{\bf G''};M^{\prime})=\Theta \left({\bf G'} \right)\Theta \left({\bf G''} \right)$ in \refeq{eq:ecplanew2} yields effective basis set energies analogous to the $E_k$-cutoff described in \refsec{sec1}.

We draw particular attention to $\chi_{i\bf G''}^{\bf G G'}(M)$, which, unlike the case of the HEG, depends implicitly on $M$ even in MP2 theory. This is due to the change in Hartree-Fock orbitals, commonly referred to as orbital relaxation, as the basis set is enlarged.

\subsection{Computational details}
We employ the Vienna {\it ab-initio} simulation package (\texttt{VASP}) in the framework of the
projector-augmented wave (PAW) method to carry out MP2 calculations of the LiH solid and molecule.\cite{Kresse1996,Kresse1999}
In the PAW method the one-electron orbitals $\psi$ are derived from
the pseudo-orbitals $\tilde{\psi}$ by means of a linear
transformation~\cite{Blochl1994}
\begin{equation}
|\psi\rangle=|\tilde{\psi}\rangle+\sum_{i}(|\phi_{i}\rangle-|\tilde{\phi}_{i}\rangle)\left\langle\tilde{p}_{i}|\tilde{\psi}
\right\rangle. 
\label{lineartransformation}
\end{equation}
The pseudo-orbitals $\tilde{\psi}$ are the variational quantities of
the PAW method, and are expanded in reciprocal space using plane waves.
We note that only the pseudo-orbitals are employed in calculating the projection matrix $U_{n\bf G}$ in \refeq{eq:ident}.
The index $i$ is a shorthand for the atomic site ${\bf R}_i$, the angular
momentum quantum numbers $l_i$ and $m_i$, and an additional index
$\epsilon_{i}$ denoting the linearization energy~\cite{Kresse1996}.
The all-electron partial waves ${\phi}_{i}$ are the solution to the radial
Schr\"odinger equation for the non-spin-polarized reference atom at
specific energies $\epsilon_i$ and specific angular momentum $l_i$.
The pseudo-partial waves, $\tilde{\phi}_{i}$, are equivalent to the
all-electron partial waves outside a core radius $r_{c}$ and match
continuously onto ${\phi}_{i}$ inside the core radius.
The partial waves $\phi_{i}$ and $\tilde{\phi}_{i}$ are represented on
radial logarithmic grids.
The projector functions $\tilde{p}_i$ are constructed in such a way that
they are dual to the pseudo partial waves, {\it i.e.},
\begin{equation}
\left\langle\tilde{p}_{i}|\tilde{\phi}_{j} \right\rangle =\delta_{ij}.
\end{equation}
For a more detailed outline of the PAW method and a thorough discussion of the
evaluation of electron repulsion integrals in \texttt{VASP}
we refer the reader to Ref.~\cite{Marsman2009}.

The employed plane wave basis set for the one-electron orbitals and the transformation
matrix $U$ is defined by all PWs $e^{i\bf{Gr}}$
with wavevectors ${\bf G}$ satisfying the equation
\begin{align*}
 (\hbar^2/2m_e)|{\bf G}|^2 < & E_{\rm cut}. \\
\end{align*}
For the calculations of LiH we use $E_{\rm cut}=400$ eV.
The evaluation of electron repulsion integrals $v_{ij}^{ab}$
in the PAW method requires an auxiliary plane wave basis set.
We choose our auxiliary plane wave basis set to be identical to
the basis set defined by $E_{\rm cut}$.

In the present work, we employ 200 and 50 natural orbitals to calculate the correlation energies of the solid and molecule, respectively. Convergence in the natural orbitals basis is two times faster than using Hartree--Fock orbitals.
Natural orbitals are calculated by diagonalizing the one-electron reduced density matrix. A detailed explanation
of this procedure can be found in Ref.~\cite{Grueneis2011}.

For the LiH solid calculations, we employ a supercell containing 8 Li and 8 H atoms.
The supercell has a volume of 136.24~\AA$^3$.
The LiH molecule is simulated using a box with a volume of 91.12~\AA$^3$ and a bond length of 1.595~\AA.
The Li 1s electrons are frozen and do not contribute to the correlation energies.



\subsection{Results: LiH molecule and solid}

\begin{figure}
\psfrag{KKK1}[l][l][1.0][0]{$E_k$ SPE} 
\psfrag{KKK2}[l][l][1.0][0]{Previous scheme} 
\psfrag{KKK3}[l][l][1.0][0]{$E_g$ SPE} 

\psfrag{AAA}[l][l][1.0][0]{CBS $\pm$ 1\%} 
\psfrag{BBB}[l][l][1.0][0]{CBS result} 
\psfrag{XXX}[][][1.0][0]{$M^{\prime -1}$} 
\psfrag{YYY}[][][1.0][0]{Correlation energy (a.u.)} 
\psfrag{X5}[][][1.0][0]{$54^{-1}$} 
\psfrag{X4}[][][1.0][0]{$117^{-1}$} 
\psfrag{X3}[l][l][0.7][0]{$337^{-1}$} 
\psfrag{X2}[l][l][0.7][0]{$1786^{-1}$} 
\psfrag{XL}[][][1.0][0]{$|$} 
\psfrag{XX}[][][1.0][0]{$\infty^{-1}$} 

\includegraphics[width=0.4\textwidth]{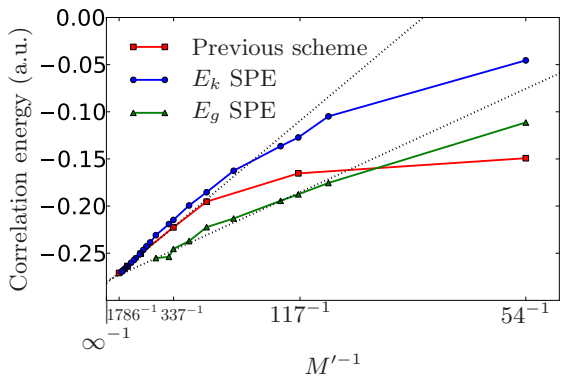}
\caption{\label{fig:lihsol}The MP2 correlation energy of the LiH 2$\times$2$\times$2 supercell
retrieved as a function of the basis set size for a variety of extrapolation schemes. The SPE curve shows the effective basis set energies produced from a single calculation with $M=2045$.}
\end{figure}

\begin{figure}
\psfrag{KKK1}[l][l][1.0][0]{$E_k$ SPE} 
\psfrag{KKK2}[l][l][1.0][0]{Previous scheme} 
\psfrag{KKK3}[l][l][1.0][0]{$E_g$ SPE} 

\psfrag{AAA}[l][l][1.0][0]{CBS $\pm$ 1\%} 
\psfrag{BBB}[l][l][1.0][0]{CBS result} 
\psfrag{XXX}[][][1.0][0]{$M^{\prime -1}$} 
\psfrag{YYY}[][][1.0][0]{Correlation energy (a.u.)} 
\psfrag{X5}[l][l][1.0][0]{$162^{-1}$} 
\psfrag{X4}[r][r][1.0][0]{$186^{-1}$} 
\psfrag{X3}[l][l][1.0][0]{$358^{-1}$} 
\psfrag{X2}[l][l][1.0][0]{$682^{-1}$} 
\psfrag{XL}[][][1.0][0]{$\infty^{-1}$} 

\includegraphics[width=0.4\textwidth,clip=true]{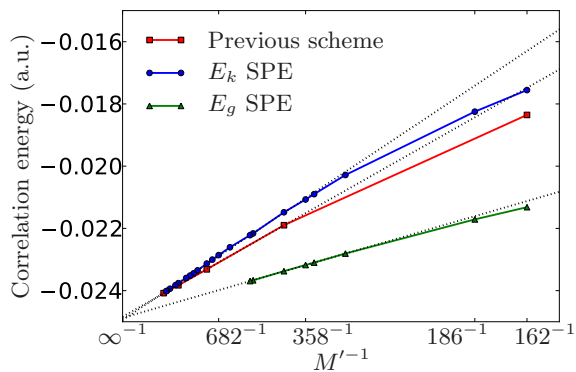}
\caption{\label{fig:lihmol}The MP2 correlation energy of the LiH molecule
retrieved as a function of the basis set size for a variety of extrapolation schemes. The SPE curve shows the effective basis set energies produced from a single calculation with $M=1647$.}
\end{figure}

In the following we will apply three different cutoff extrapolation schemes to the LiH solid and molecule using MP2:
(i) the local $E_g$ cutoff extrapolation scheme that is equivalent to the one previously outlined in Ref.~\onlinecite{Marsman2009}
(ii) the union $E_g$ cutoff, and (iii) the $E_k$-cutoff.
Figures \ref{fig:lihsol} and \ref{fig:lihmol} show the convergence of the MP2 correlation energy of the LiH solid
and molecule, respectively. Both single-point extrapolations show a much-improved behaviour over the previous scheme that is analogous to a local $E_g$-cutoff, where arcing causes pathological behaviour and poor CBS estimates at low $M^{\prime}$.
In both solid and molecular LiH, the (union, SPE) $E_g$-cutoff seems to converge quicker. 

%
%

\section{Concluding remarks}

In this paper, we have investigated the convergence of correlation energies using plane-wave wave-function expansions. Starting by treating the finite simulation-cell electron gas with the simplest correlated quantum chemical method, second-order M\o ller-Plesset perturbation theory, we derive a functional form of the finite basis set correlation energy of $1/M$, where $M$ is the number of plane waves enclosed by a spherical cutoff in $k$-space. Although perturbation theory diverges in metallic systems for any strength of Coulomb interaction, the qualitative behaviour of the wavefunction around the correlation hole is in common with other higher-level methods. We verify that  this $1/M$ behaviour extends to coupled-cluster doubles (CCD) and the random-phase approximation plus second-order screen exchange (RPA+SOSEX), in common with exact results from full configuration interaction quantum Monte Carlo (FCIQMC)\cite{UEGPaper1, UEGPaper2preprint}. 

By viewing the distribution of the wavefunction in configuration space over double-excitations, and relating this to orbital momenta in $k$-space, we propose several new basis set truncations based on the momentum transfer vector. We discuss these in terms of their comparative speed and smoothness of convergence recommending one scheme, which we call the \emph{union} $E_g$-cutoff, that gives overall the most desirable properties. This is then generalised to other single-reference quantum chemical techniques, allowing for the development of a \emph{single-point extrapolation} technique which uses information from a single large-basis-set calculation to provide estimates for the complete basis set limit correlation energy in CCD and RPA+SOSEX.

Finally, this is applied to real materials (molecular and solid LiH). We find that the energies computed by single-point extrapolation converge better and more reliably than previous extrapolation techniques\cite{Marsman2009}. It is our hope that this can be applied in future plane-wave wavefunction based calculations.

\acknowledgments

This work was supported by a grant from the Distributed European Infrastructure for Supercomputing Applications under their Extreme Computing Initiative and EPSRC (JJS, AA) for funding. One of us (GK) acknowledges support of the Austrian Science Fund (FWF) within the SFB ViCoM (F41).

\begin{appendix}
\section{CCD and RPA+SOSEX}
\label{sec:ccd}
In the following, we will briefly outline coupled-cluster doubles (CCD) theory and the
random-phase approximation plus second-order screened exchange (RPA+SOSEX).

CCD is a widely used quantum chemical method to study the electronic ground state energy of atoms and molecules and relies on
an exponential Ansatz for the many-electron wavefunction that reads \cite{Bartlett2007,Cizek1966}
\begin{equation}
\Psi^{\rm CCD}=e^{\hat{T}_2} \Psi_0,
\end{equation}
where $\hat{T}_2$ refers to the double excitation operator.\cite{Bartlett2007}
\begin{equation}
{{\hat{T}}_2} |\Psi_{0} \rangle= \sum_{i<j}^{\rm occ.} \sum_{a<b}^{\rm unocc.}
t_{ij}^{ab} p^\dagger_a p^\dagger_b p_i p_j |\Psi_{0} \rangle=\sum_{i<j}^{\rm occ.} \sum_{a<b}^{\rm unocc.} t_{ij}^{ab} | \Psi_{ij}^{ab} \rangle.
\end{equation}
We choose $\Psi_0$ to be the Hartree--Fock reference determinant.
The solution to the CCD wavefunction is obtained by projecting $\Psi^{\rm CCD}$ onto a set of doubly
excited determinants.
This set of equations is termed amplitude equations. 
The CCD amplitude equations read~\cite{Forner1997}
\begin{equation}
\begin{split}
0=&v_{ij}^{ab}+(\epsilon_a^{\rm HF}+\epsilon_b^{\rm HF}-\epsilon_i^{\rm HF}-\epsilon_j^{\rm HF})t_{ij}^{ab}
\\&+\sum_{lc}
[(2v_{ic}^{al}-v_{ci}^{al})t_{lj}^{cb}-v_{ic}^{al} t_{lj}^{bc}-v_{ci}^{bl} t_{lj}^{ac}
\\&\quad+(2v_{jc}^{bl}-v_{cj}^{bl})t_{li}^{ca}-v_{jc}^{bl} t_{li}^{ac}-v_{cj}^{al} t_{li}^{bc}]
\\&+\sum_{cc'}v_{cc'}^{ab} t_{ij}^{cc'}+\sum_{ll'}v_{ij}^{ll'} t_{ll'}^{ab}
\\&+\sum_{ll'}\sum_{cc'}
[(2v_{cc'}^{ll'}-v_{c'c}^{ll'})
(
2t_{il}^{ac}t_{jl'}^{bc'}-t_{li}^{at}t_{jl'}^{bc'}
\\&-t_{il}^{ac}t_{l'j}^{bc'}-t_{li}^{cc'}t_{l'j}^{ab}-t_{lj}^{cc'}t_{l'i}^{ba}-t_{l'l}^{ac}t_{ji}^{bc'}
\\& -t_{l'l}^{bc}t_{ij}^{ac'})
+v_{cc'}^{ll'}(t_{li}^{at}t_{l'j}^{bc'}+t_{lj}^{ac'}t_{l'i}^{bt}+t_{ll'}^{ab}t_{ij}^{cc'})
]
\end{split}
\label{eq:Tccd}
\end{equation}
$i,j,l$ and $a,b,c$ refer to occupied and unoccupied orbitals, respectively.
The amplitude equations can also be written in a more compact fashion by defining intermediate quantities\cite{Hirata2001}. 
Solving Eq.~(\ref{eq:Tccd}) yields the wavefunction coefficients in configuration space $t_{ij}^{ab}$ and allows for the
correlation energy to be calculated according to Eq.~(\ref{eq:corren}).
Due to the computational cost involved, CCD has so far only rarely been applied to solid state systems.

Freeman, and Bishop and L\"uhrmann studied the uniform electron gas using an approximation to CCD theory.\cite{Freeman1977,Bishop1978}
This approximation has recently attracted renewed interest and is termed RPA+SOSEX.\cite{Angyan2011,Toulouse2011,Klopper2011,Hesselmann2011,Jansen2010,Paier2010,Grueneis2009}
RPA+SOSEX differs from CCD in two points:
(i) the HF reference is replaced by the KS reference, which greatly reduces the one-electron gap
and, (ii) the double amplitude equations are approximated by so-called ring diagrams only
\begin{eqnarray}
0& =& v_{ij}^{ab} +t_{ij}^{ab}(\epsilon^{KS}_a+\epsilon^{KS}_b-\epsilon^{KS}_i-\epsilon^{KS}_j)
\nonumber \\
&+ &\sum_{lc} v_{ic}^{al} t_{lj}^{tb} + \sum_{lc} t_{il}^{ac} v_{cj}^{lb} + \sum_{ll'cc'} t_{il}^{ac} v_{cc'}^{ll'} t_{l'j}^{c'b},
\label{eq:Tsosex}
\end{eqnarray}
Once obtained, the RPA+SOSEX $t_{ij}^{ab}$-amplitudes can be employed to calculate the RPA+SOSEX correlation energy using Eq.~(\ref{eq:corren}).
A rigorous justification for this approximation is not straightforward and would be beyond the scope of this work.
However, Ref.~\onlinecite{Scuseria2008} outlines the connection between the above amplitude and Casida's equation.

\end{appendix}


\end{document}